\documentclass[trackchanges,twocolumn]{aastex701}

\usepackage{newtxtext,newtxmath}
\usepackage[T1]{fontenc}
\usepackage{lineno}
\usepackage{graphicx}
\usepackage{amsmath}
\usepackage{hyperref}
\usepackage{longtable}
\usepackage[figuresright]{rotating}
\usepackage{multirow}
\usepackage{afterpage}
\usepackage{enumitem}
\usepackage{placeins}

\begin{document}

\title{A Spectral Framework for Testing the Quasi-Star Hypothesis in Little Red Dots I: Weighing LRDs by Their Super-Eddington Luminosity Ratios---No Signs of Overmassive Black Holes}

\author[0000-0002-0212-4563]{Olivia Curtis}
\affiliation{Department of Astronomy and Astrophysics, The Pennsylvania State University, 251 Pollock Road, University Park, PA 16802, USA}
\affiliation{Institute for Gravitation and the Cosmos, The Pennsylvania State University, University Park, PA 16802, USA}
\affiliation{Penn State Extraterrestrial Intelligence Center, 525 Davey Laboratory, 251 Pollock Road, Penn State, University Park, PA, 16802, USA}
\email[show]{ocurtis@psu.edu}

\author[0000-0001-7151-009X]{Nikko J. Cleri}
\affiliation{Department of Astronomy and Astrophysics, The Pennsylvania State University, 251 Pollock Road, University Park, PA 16802, USA}
\affiliation{Institute for Computational and Data Sciences, The Pennsylvania State University, University Park, PA 16802, USA}
\affiliation{Institute for Gravitation and the Cosmos, The Pennsylvania State University, University Park, PA 16802, USA}
\email{cleri@psu.edu}

\author[0000-0003-4337-6211]{Jakob M. Helton}
\affiliation{Department of Astronomy and Astrophysics, The Pennsylvania State University, 251 Pollock Road, University Park, PA 16802, USA}
\email{jakobhelton@psu.edu}

\author[0000-0001-6755-1315]{Joel Leja}
\affiliation{Department of Astronomy and Astrophysics, The Pennsylvania State University, 251 Pollock Road, University Park, PA 16802, USA} 
\affiliation{Institute for Computational and Data Sciences, The Pennsylvania State University, University Park, PA 16802, USA}
\affiliation{Institute for Gravitation and the Cosmos, The Pennsylvania State University, University Park, PA 16802, USA}
\email{joel.leja@psu.edu}

\author[0000-0001-6160-5888]{Jason T.\ Wright}
\affiliation{Department of Astronomy and Astrophysics, The Pennsylvania State University, 251 Pollock Road, University Park, PA 16802, USA}
\affiliation{Center for Exoplanets and Habitable Worlds, 525 Davey Laboratory, 251 Pollock Road, Penn State, University Park, PA, 16802, USA}
\affiliation{Penn State Extraterrestrial Intelligence Center, 525 Davey Laboratory, 251 Pollock Road, Penn State, University Park, PA, 16802, USA}
\email{astrowright@gmail.com}

\begin{abstract}

We present a spectral test of the quasi-star hypothesis for Little Red Dots (LRDs) whereby a black hole grows inside a stellar-like envelope. We use \texttt{Prospector} to fit host galaxies and \texttt{TLUSTY} photospheres to 5 LRDs that show strong molecular or atomic absorption. We approximate each object's electron-scattered Eddington luminosity ratio, $\phi \equiv \kappa_{\rm es}\sigma T_{\rm eff}^{4}/(gc)$, and we use \texttt{MESA-QUEST} to simulate their envelopes. All 5 are super-Eddington at $\phi = {87}$--$299$, with the caveat that the largest sources sit at the edge of our atmosphere grid and beyond our simulations. We derive envelope masses between $700$--${15{,}000}\,M_\odot$, where quasi-star theory requires the black hole to be less than a third of that. GN-28074 falls 5 decades below its published virial mass estimate, alleviating the overmassive black hole problem. Their black holes double every $\sim0.05$--${0.2}$~Myr and can produce intermediate-mass black holes in $\lesssim30$~Myr. Our 3 water-absorbing objects have cold components that are $2$--$3$~dex denser than their hot components, which we interpret as the water forming in cold dense clouds. We then extend our measurements to 82 archival LRDs, finding that the population has super-Eddington photospheres and wind speeds that increase with $\phi$, which implies that LRDs evolve from massive sources with slow winds to having more eruptive winds as they mature and shed their outer envelopes. We thus constrain the Eddington ratios and masses of LRDs and show self-consistently that quasi-stars may be the central engines powering LRDs.

\end{abstract}

\keywords{Active galactic nuclei (16) --- Supermassive black holes (1663) --- Black hole physics (159) --- Stellar atmospheres (1584) --- High-redshift galaxies (734)}

\section{Introduction}
\label{sec:intro}

The discovery of ``Little Red Dots'' (LRDs, e.g., \citealt{Matthee2024, Greene2024, deGraaff2025, Hviding2025}) is one of the most definitive achievements of the James Webb Space Telescope (JWST). The population was identified within the first year of operations \citep{Kocevski2023, Labbe2025} and has since been observed across several large photometric and spectroscopic surveys \citep{Kokorev2024, Kocevski2025, Weibel2026}. Even so, all aspects of their nature, namely their broad emission lines and breaks, their v-shaped UV-optical spectrum, and their lack of X-ray emission, remain challenging for any single model to describe. Many proposed explanations describe some form of black hole star \citep{deGraaff2025,Naidu2025}, i.e., a black hole enshrouded in gas that exhibits the classical features of a black hole (e.g., broad lines), together with those of a star (e.g., a Balmer break). Edge cases also compound the problem of producing a unified theory by disrupting each selection criterion as it emerges. Among them are the X-ray detections of a minority of otherwise X-ray-quiet sources \citep[e.g.,][]{Fu2025, Hviding2026xrd}, the emerging long-baseline variability of individual objects \citep{Cantiello2025, Zhang2025}, and the strong absorption features that we study below (e.g., \citealt{Lin2025, Wang2026}).

If a unified theory for LRDs exists, then it must account for each of these edge cases, and several of these proposed models are not mutually exclusive. Many build on the observation that the optical to near-infrared continua of LRDs are well-fit by an optically thick stellar-photosphere-like blackbody \citep{deGraaff2025, Sun2026, Umeda2026}, although some studies (e.g., \citealt{Sneppen2026paschen}) report Paschen jumps in some LRDs that are more consistent with nebular rather than photospheric origins. At surface gravities $\log g<0$, \citet{Begelman2026}, \citet{Gentile2026}, and \citet{Liu2026} showed that such photospheres naturally reproduce several LRD observables, including their temperatures \citep{deGraaff2025}, red colors \citep{Kido2025}, strong Balmer lines \citep{Ji2025, Taylor2025}, broad electron-scattered Balmer lines \citep{Rusakov2026, Sneppen2026cocoon, Matthee2026}, and Balmer absorption \citep{Matthee2024, Yanagisawa2026, Juodzbalis2026}.

The direct-collapse black hole literature, which in our usage refers to scenarios whereby a massive seed forms when a metal-poor gas cloud collapses to form a supermassive star and later a massive black hole seed, provides several scenarios that produce optically thick photospheres around accreting seed black holes (e.g., \citealt{Begelman2006}; \citealt{Pezzulli2016}; \citealt{Luo2020}; \citealt{RomanGarza2026}). The most promising, in part because of its $1$--$20\,$Myr lifetime \citep{Ball2012,Santarelli2026}, is the quasi-star, which forms when a gas reservoir of $10^{4}$--$10^{6}\,M_\odot$, assembled through the direct collapse of a halo or a dense filament, collapses into a supermassive star whose core collapses to form a stellar-mass black hole surrounded by a bound stellar envelope \citep{Begelman2006, Begelman2008}. But supermassive stars can also be formed in runaway stellar collisions in dense star clusters (e.g., \citealt{PortegiesZwart2002, Fujii2024, Pacucci2025, Rantala2026}), or in already enriched gas \citep{ChonOmukai2020} that is being externally heated through Lyman-Werner radiation \citep{Bromm2003, Regan2016}.

In this manuscript, we do not concern ourselves with how the supermassive star forms and instead focus on what happens to the star and its environment after the core of the supermassive star collapses and leaves behind a black hole embedded in a radiation-supported, fully convective envelope. Because the envelope reprocesses the accretion luminosity, its luminosity is governed by the Eddington limit of the entire object such that the black hole is able to grow by 3 to 4 orders of magnitude within a few Myr while its photosphere resembles a cool, extremely low-gravity supergiant \citep{Ball2011, Coughlin2024, Hassan2025}. The inner-most regions of these objects, the so-called saturated convection layers of \citet{Coughlin2024}, have analytic solutions that limit the black hole mass and provide a transport mechanism that moves the energy generated by accretion into the outer envelope. The saturated convection layer thus acts as the ``core'' of the quasi-star, acting as the energy transport layer that supports the layers above it (see Figure~\ref{fig:cartoon}). 

Now, the broader term black hole star (e.g., \citealt{Naidu2025, deGraaff2025, Sun2026}) is agnostic to the formation channel and applies to any black hole surrounded by a dense gas envelope, photosphere, or atmosphere. A quasi-star is not the only way to surround a black hole with a dense gaseous envelope; for instance, failed winds from super-Eddington accretion flows \citep{Liu2025,Lambrides2026,LiuFengHo2026} or a supermassive star that collapses directly to a $\sim10^{4}\,M_\odot$ black hole \citep{Nandal2026} also produce black-hole-star-like spectra. In this paper, we place a quasi-star inside an extended gaseous atmosphere to explain the molecular and atomic absorption features that some sources exhibit, and throughout this manuscript, we refer to our model as a quasi-star to differentiate it from other black hole star models.

There has recently been a growing interest on the theory behind quasi-stars. For example, \citet{Campbell2025} and \citet{Santarelli2026} have developed the \texttt{MESA-QUEST} framework, which simulates the evolution of the saturated convection layer and its outer layers using the Modules for Experiments in Stellar Astrophysics (\texttt{MESA}) code (e.g., \citealt{Paxton2011,Paxton2013, Paxton2015, Paxton2018,Paxton2019,Jermyn2023}), where it successfully predicts the photospheric temperatures and lifetimes of LRDs. \citet{Hassan2025} independently implement quasi-stars into \texttt{MESA}, where they find that their envelopes can remain stable until $M_{\rm BH}/M_{\rm rad} \approx 0.33$, after which the quasi-star destabilizes and becomes a typical active galactic nucleus. We adopt that black hole mass to envelope mass relation throughout, where we use $M_{\rm BH}$ to denote the mass of the black hole, $M_{\rm rad}$ is the mass interior to our simulated radiative layers (see Figure~\ref{fig:cartoon}), and $M_\bigstar$ is the mass of everything beneath the extended atmosphere (i.e., what we actually observe). We also use $M_*$ to indicate the total stellar mass of the host galaxy. 

Although these models seem to indicate that quasi-stars can remain stable for $\sim1$--$20$~Myr, \citet{Santarelli2026} and \citet{Hassan2025} have argued that continuous accretion from the birth cloud onto the stellar envelope can counteract wind losses, thus extending the lifetime of the quasi-star, and, indeed, \citet{RomanGarza2026} evolve such accreting quasi-stars to lifetimes that can reach $10$--$100$~Myr. This agrees with the results of \citet{Sun2026}, who measure a duty cycle near 1$\%$ from the population's number density of $\approx10^{-5}\,\rm{cMpc}^{-3}$ and conclude that every supermassive black hole may have spent $\sim10$~Myr in an LRD phase. \citet{Begelman2026} further interpret LRDs as the late stage of this evolution, and \citet{Cantiello2025} show that these cool envelopes can even cross a fundamental-mode instability strip that may explain the emerging variability of the population \citep{Zhang2025}. Lastly, \citet{Gentile2026} have fit blackbody quasi-star models to 86 LRD spectra and reproduce the continuum shapes, Balmer breaks, and hydrogen-line luminosities of the population.

Recently, \citet{Naidu2026} argue that LRD continua, Balmer profiles, and molecular features closely resemble the wind photospheres of $\eta$ Carinae's Great Eruption and Type IIn supernovae. From the Eddington limit, the blueshifted Balmer absorption, the absence of rapid variability, and the low fitted gravities, they bound the black hole masses below $10^{5.1}\,M_\odot$ and conclude that their sources are super-Eddington with electron-scattering Eddington luminosity ratios $\Gamma_{\rm es}\approx5$--$50$. In their picture, similar to the quasi-star case (e.g., \citealt{Begelman2026}), the broad Balmer lines form from electron-scattering within the low surface-gravity winds such that their widths contain no virial information.

Much like the case for stars, only the (pseudo-)photospheres of LRDs are directly observable, and several authors have tried to explain their observed phenomena. Blackbody continua fit the population's optical colors and shape \citep{deGraaff2025, Gentile2026}, photospheric Balmer absorption features are seemingly ubiquitous across LRDs \citep{Matthee2024, Yanagisawa2026, Juodzbalis2026}, and cool, low-gravity stellar atmospheres reproduce the strongest absorption features \citep{Lin2025, Liu2026, Wang2026}. We extend that and ask, if LRDs or quasi-stars have photospheres, then are there other components of stellar atmospheres that describe the upper atmospheres of these objects? Below, we describe several such structures that can reproduce the observed features of LRDs while remaining consistent with the energy budget of a quasi-star, and much of this manuscript is thus dedicated to linking these two regimes. For instance, we describe a tiered atmosphere of inflowing and outflowing material that produces Balmer progressions that are similar to the types seen in T Tauri stars \citep{Walker1972} or the equatorial inflow--polar-outflow geometry that \citet{Matthee2026} propose and that \citet{Sneppen2026cocoon} infer from their radiative-transfer models; we place a chromosphere above that layer, which can supply enough ionizing photons to produce the saturated \ion{He}{1} $\lambda10830$ troughs that the data already show (e.g., The Rosetta Stone, \citealt{Juodzbalis2024}; GLIMPSE-17775, \citealt{Kokorev2026}; MoM-BH$^{*}$-1, \citealt{Naidu2025}; RUBIES-BLAGN-1, \citealt{Wang2025}), and we discuss how even modest magnetic fields in the photosphere and corona can confine starspots on the surface (see below and Paper~II in preparation). 

In this manuscript, we derive super-Eddington luminosity ratios, and thus envelope masses, on an object-by-object basis, and, in doing so, provide a framework that can test the quasi-star hypothesis as an explanation for the central engines of LRDs. We introduce a mass-independent Eddington ratio, $\phi$, constructed purely from fitted photospheric parameters, and we measure it for the 5 LRDs that show the strongest molecular and atomic absorption features as well as for 82 other archival sources. We thereby provide an object-by-object demonstration that $\Gamma_{\rm es} > 1$ for most LRDs. We use our fitted $\phi$ values to infer accretion factors, wind properties, envelope masses, and black hole masses, and, in doing so, test the eruptive outflow model of \cite{Naidu2026}. We thus show that several LRD observables are fully consistent with the energy budget provided by quasi-stars \citep{Coughlin2024, Hassan2025, Santarelli2026, Begelman2026}.

Section~\ref{sec:methods} presents the data and our spectral fitting code. In Section~\ref{sec:mesaquest}, we develop the quasi-star model and our $\phi$ parameter. Section~\ref{sec:results} shows our photospheric fits and our derived masses. Section~\ref{sec:discussion} then compares masses and wind speeds across our fitted sample, and we also hypothesize on an evolutionary sequence for LRDs based on our results. We summarize our conclusions in Section~\ref{sec:conclusion}, and we describe the interiors of simulated quasi-stars as well as higher-order corrections to our mass measurements in the appendices. Throughout, we assume a flat $\Lambda$CDM cosmology with $H_0 = 69.3$~km\,s$^{-1}$\,Mpc$^{-1}$ and $\Omega_{\rm m} = 0.287$ \citep{Hinshaw2013}.

\section{Observations and Atmosphere Models}
\label{sec:methods}

\subsection{Observations}
\label{sec:obs}

Our primary analysis uses a sample of 5 LRDs that show very strong molecular and atomic absorption features, as well as an archival sample of $82$ LRDs, most of which have broad Balmer features. Our main sample of 5 galaxies is comprised of J1025+1402 (redshift $z=0.1007$, hereafter ``The Egg,'' \citealt{Lin2025}) which shows a very strong \ion{Ca}{2} triplet, GN-28074 ($z=2.26$, ``The Rosetta Stone,'' \citealt{Juodzbalis2024}) which shows a tentative detection of a \ion{Ca}{2} triplet, and WIDE-EGS-2974 ($z=2.3203$), UNCOVER-A2744-20698 ($z=2.42$) and CAPERS-UDS-23216 ($z=2.2957$; \citealt{Wang2026}) which all show signs of molecular water absorption. Going forward, we will collectively term the latter 3 sources as ``water dots'' after their signature spectral feature. 

We use JWST NIRSpec PRISM observations of 3 water dots, taking the reduced 1D spectra from the DAWN JWST Archive \citep{Heintz2025, deGraaff2025rubies}. WIDE-EGS-2974 lies in the Extended Groth Strip and comes from the NIRSpec WIDE GTO Survey (GTO 1213, PI N.~Luetzgendorf; \citealt{Maseda2024}). UNCOVER-A2744-20698 sits behind the Abell 2744 cluster field and comes from the Ultradeep NIRSpec and NIRCam ObserVations before the Epoch of Reionization survey (UNCOVER; GO 2561, PI I.~Labb\'e; \citealt{Bezanson2024}). CAPERS-UDS-23216 is in the Ultra Deep Survey (UDS) field of the United Kingdom Infra-Red Telescope Infrared Deep Sky Survey \citep{Lawrence2007}, as cataloged by the CANDELS-Area Prism Epoch of Reionization Survey (CAPERS; GO 6368, PI M.~Dickinson; \citealt{Kokorev2025capers}).

The Egg is a local LRD from the Sloan Digital Sky Survey \citep{SDSS}. \cite{Lin2025} identified it as a low-redshift analog of the high-$z$ LRD population and observed it with the Large Binocular Telescope (LBT) Multi-Object Double Spectrograph (MODS; $3300$--$10{,}000$~\AA, $R\sim800$--$1700$) as well as with the Magellan Folded-port InfraRed Echellette (FIRE; $0.8$--$2.5\,\mu$m) spectrometer. The Egg shows very high equivalent width \ion{Ca}{2} absorption, along with weaker low-ionization metal features characteristic of cool stellar photospheres \citep{Lin2025}. \cite{Liu2026} fit these observations with a suite of low surface gravity ($\log g<0$) \texttt{TLUSTY} \citep{Hubeny1988, Hubeny2021} atmosphere models and found that The Egg requires an effective temperature $T_{\rm eff}=4500$~K and surface gravity $\log g=-2.90$ to explain its strong absorption features (Section~\ref{sec:TLUSTY}).

The Rosetta Stone \citep{Juodzbalis2024} lies in the Great Observatories Origins Deep Survey North (GOODS-N) field and comes from the JWST Advanced Deep Extragalactic Survey (JADES; GTO 1181, PI D.~Eisenstein; \citealt{Eisenstein2023}). We use its JWST NIRSpec PRISM observations as well as its 3 medium-resolution gratings (G140M, G235M, and G395M, all at $R\sim1000$). Its continuum displays the characteristic v-shape and red optical colors, and, if its broad lines are interpretted as being Doppler broadened, it has a virial estimate of $\log(M_{\rm BH}/M_\odot)\approx8.5$ \citep{Juodzbalis2024}. Much like The Egg, The Rosetta Stone shows signs of a \ion{Ca}{2} triplet.
  
Our broader sample consists of 82 supplementary LRDs, which we draw from four archival catalogs and 5 individually published sources. We primarily select these sources for the wind analysis that we perform in Section~\ref{sec:disc_popwinds}, where we assemble a large population of sources that have a published Balmer absorption to test how terminal wind speeds vary as a function of our fitted $\phi$ values. We use 11 sources compiled by \citet{Naidu2026}, 30 sources compiled by \citet{Hviding2025}, 8 used in the study by \citet{Sok2026}, and 28 from the low-redshift DESI sample (e.g., \citealt{Park2026,Lin2026}). The remaining 5 are A2744-QSO1 \citep{Ji2025, DEugenio2026}, GLIMPSE-17775 \citep{Kokorev2026}, MoM-BH$^{*}$-1 \citep{Naidu2025}, and CEERS-6126 and UDS-31092 \citep{Kocevski2025}. We restrict our low-redshift DESI sources to the spectroscopically vetted (LRDs)$^2$ sample \citep{Lin2026}. We take the JWST spectra from the DAWN JWST Archive and the DESI spectra from the DESI Data Release 1 \citep{DESI2026}, and we fit each at its native resolution with the framework of Section~\ref{sec:prospector}.

\subsection{TLUSTY Atmospheres}
\label{sec:TLUSTY}

We model the photospheric emission with the synthetic spectral library of \citet{Liu2026}, which is computed with the plane-parallel radiative-transfer code \texttt{TLUSTY} \citep{Hubeny1988, Hubeny2021}. Conventional stellar atmosphere grids terminate near $\log g \approx 0$, whereas LRD continua require gravities several dex lower than that, so \citet{Liu2026} extended the calculations into the radiation-pressure-dominated regime where the atmosphere approaches the local Eddington limit. The library contains 233 models that span $T_{\rm eff} = 2000$--$7500$~K, $\log g = -4$ to $+1.5$, metallicities $[{\rm M/H}] = -2$, $-1$, and $0$, and it includes models with microturbulent velocities of 2 and 10~km\,s$^{-1}$. Only the $[{\rm M/H}] = -1$ grid covers temperatures lower than $\sim4,000$K, so we fix the photospheric metallicities of the water dots to this metallicity since water is not stable at temperatures higher than $\sim3,000$K. While sampling, we bracket the sampled temperatures and gravities by the two nearest grid values and interpolate linearly between the two nodes. The models are hydrostatic and plane parallel, both of which affect all of our measurements (see below).

\subsubsection{Consequences of Modeling in Hydrostatic Equilibrium}
\label{sec:consequences}

An important caveat to this study is the fact that our models assume the atmosphere is plane-parallel and in hydrostatic equilibrium (see, e.g., \citealt{Lebzelter2012, Hubeny2021, Gonzalez-Tora2023, Liu2026}). \citet{Liu2026} directly quantify the assumption of hydrostatic equilibrium while creating their models, and we adopt their formalism since our $\phi$ values depend on it (Section~\ref{sec:phi}). A spectrum does not measure the local acceleration due to gravity, $g$, directly since it is the pressure and temperature that set the ionization balance, the electron pressure, and the opacities. A closure relation is then needed to convert that pressure into a gravity. Hydrostatic equilibrium can provide that closure, but that does not hold for an outflow layer that is accelerating. \citet{Liu2026} retain the dynamical term in the radial equation of motion, so the parameter that their library constrains, and what we actually fit, is the net effective gravity $g_{\rm net} \equiv g - g_{\rm dyn}$. Here, $g = GM_{\rm tot}/R^{2}$ is the true local gravity for a gravitational constant $G$, a mass $M_{\rm tot}$ that is enclosed by the photosphere, and a photospheric radius $R$, while $g_{\rm dyn} \simeq -{\rm sgn}(dv^{2}/dR)\,v^{2}/R$ is a catchall for any inward or outward motion of that layer at a radial velocity $v$, where $\rm{sgn}$ refers to the signum function. Their photospheric density is $\rho_{\rm ph} \approx 2\mu m_{p}(g_{\rm net} - g_{\rm rad})/(3 k_{B} T_{\rm eff} \kappa_{\rm ph})$, where $\mu$ is the mean molecular weight, $m_{p}$ is the proton mass, $k_{B}$ is the Boltzmann constant, and $\kappa_{\rm ph}$ is the photospheric opacity. The radiative acceleration, $g_{\rm rad} = \kappa_{F}\sigma T_{\rm eff}^{4}/c$, is derived from the flux-mean opacity $\kappa_{F}$, the Stefan--Boltzmann constant $\sigma$, and the speed of light $c$.

\citet{Liu2026} then interpret the atmospheres that approach the Eddington limit in two ways. In the first case, the inferred gravity is moderately low, so the outer layers stay sub-Eddington and hydrostatic while the inner layers pass the local Eddington limit. \citet{Liu2026} hold the gas pressure at its maximum value through those inner layers since a gas-pressure inversion is unstable and the inner atmosphere can smear it out \citep{Jiang2015}, and they show that the modification leaves the synthetic spectra largely unchanged since the outer layers, where most spectral features form, end up remaining in hydrostatic equilibrium. In the second case, the gravity is low enough that the whole atmosphere passes the Eddington limit, so \citet{Liu2026} then remove the radiative acceleration from the hydrostatic equation and interpret the pseudo-photosphere as an optically thick gas layer rather than as an atmosphere. In Appendix~\ref{app:gamma}, we show that the internal structure of our quasi-star simulations more closely aligns with the first case.


\subsection{Spectral Modeling}
  \label{sec:prospector}

We fit every source with a single forward model that we build with \texttt{Prospector} \citep{Johnson2021}.\footnote{Our \texttt{Prospector} fork, which implements the TLUSTY photosphere basis alongside the code's stock stellar population sources, is available at \url{https://github.com/o-curtis/prospector-tlusty}.} Our model simultaneously fits one or more TLUSTY photospheres with a Flexible Stellar Population Synthesis (\texttt{FSPS}) \citep{Conroy2009} host galaxy that we generate using a six-bin continuity star-formation history, a \citet{Kroupa2001} initial mass function, and a free stellar metallicity parameter. We then also add \citet{DraineLi2007} dust emission with its minimum radiation-field intensity $U_{\rm min}$, polycyclic aromatic hydrocarbon mass fraction $q_{\rm PAH}$, and we leave the high-intensity photon fraction $\gamma$ left free. We attenuate both the galaxy and photosphere's light with a shared foreground dust screen. Each \texttt{TLUSTY} photosphere has its own temperature, surface gravity, and luminosity terms. After performing a series of tests to find which dust laws best fit our sources, we adopt a Small Magellanic Cloud attenuation law \citep{Gordon2003} for every source in this work. All sampling is performed with dynamic nested sampling (\texttt{dynesty}; \citealt{Speagle2020}), which also returns the Bayesian evidence $\mathcal{Z}$, i.e., the likelihood integrated over the prior, which we use to compare model variants through their $\log_e$ evidence differences $\Delta\ln\mathcal{Z}$.

We fit every spectrum in its native pixels, where we add a systematic floor in quadrature to the pipeline uncertainties, $5\%$ for The Egg and $10\%$ for the JWST sources, where The Egg gets a lower systematic uncertainty since its MODS-R spectra constrain the continuum to $\lesssim1\%$. For The Egg, we scale its FIRE fluxes and uncertainties by a fixed factor of $1.106$ since its FIRE spectrum is $\sim9.6\%$ fainter than its MODS-R spectrum in the $7550$ to $8450$~\AA\ region. We forward-model each arm at $R = 1850$ and $2250$ for the MODS G400L and G670L spectra, $R = 6000$ for the FIRE spectra, $R = 100$ for NIRSpec PRISM spectra, $R = 1000$ for the NIRSpec medium gratings, and $R = 2500$, $3500$, and $4500$ for the DESI B, R, and Z spectra. We fit UNCOVER-A2744-20698 at $R = 300$, which is the PRISM spectral resolution where its water band falls. 

We then mask the emission lines before fitting using a single emission line list that masks the Balmer and Paschen series, \ion{He}{1} $\lambda10830$, \ion{He}{2} $\lambda4686$, [\ion{O}{3}] $\lambda\lambda4959,5007$, [\ion{Ne}{3}] $\lambda3869$, [\ion{O}{2}] $\lambda3727$, \ion{Mg}{2} $\lambda2798$, and \ion{C}{2}] $\lambda2326$. For our 5 primary sources, we modified the list to remove only lines that were actually present in the data, but, for our 82 archival fits, this full list was used. We are justified in masking the emission lines since our photospheric models are in LTE and contain no emission features \citep{Liu2026}. Models in LTE are most accurate in the continuum and absorption \citep{Hubeny1988,Hubeny2021}, while a robust forward model of the emission lines would require accounting for the host narrow lines, the electron-scattered broad lines \citep{Greene2024, Hviding2025}, lines broadened in the winds, absorption \citep{Rusakov2026, Naidu2026}, as well as any narrow coronal or chromosphere emission (e.g., \citealt{Gentile2026}, Paper~II), which is beyond the scope of this study.

We also mask the Balmer break from our fits, which we define as the region between $3400$--$4100$~\AA. As \citet{Liu2026} describe, their plane-parallel assumption fails to fully model the break, especially near the \ion{Ca}{2} H and K line cores, and they state that this assumption is likely to break down for the cold, dense, extended atmospheres like the ones that we propose in the following section. Our emission line masking already masks most of our PRISM data in this region, whereas our DESI sources resolve individual lines in the region. Independent testing has shown that, while leaving this region unmasked does not change our population-level results, the DESI sources can have their fitted $\log g$ values inflated by $\sim0.6$ dex if the region is left unmasked.

\begin{figure}
\centering
\includegraphics[width=\columnwidth]{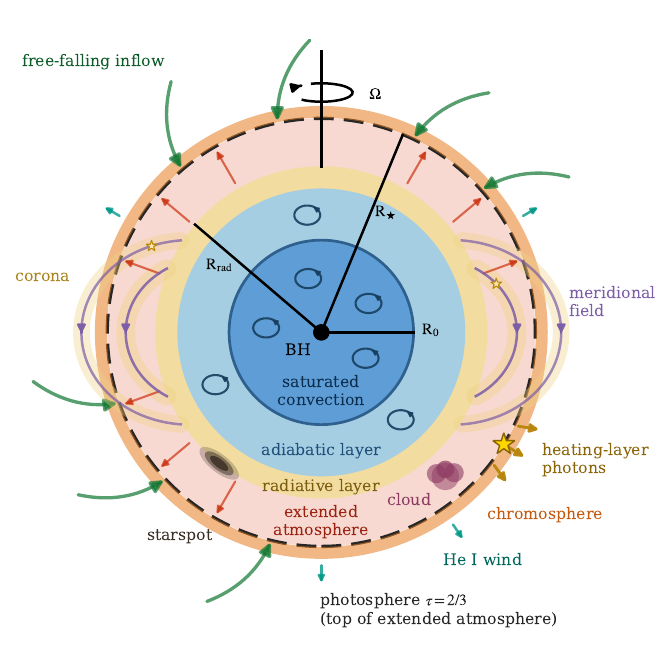}
    \caption{A cartoon of our quasi-star configuration, drawn in the meridional plane at a late evolutionary stage and not to scale. Each labeled structure is either an analytic part of the quasi-star model (e.g., the saturated convection layer), is solved by \texttt{MESA-QUEST} (the adiabatic, and radiative layers), is motivated by an observation in this paper (the extended atmosphere, and the inflow--outflow interface), is a plausible explanation for one of our observations (starspots and clouds), or is a feature that will be discussed in Paper~II (the chromosphere, the corona, and the \ion{He}{1} wind). A black hole sits at the center of the saturated-convection zone, which is surrounded by an adiabatic layer, a physically thin radiative layer, and an optically thick extended atmosphere. The dashed circle is the $\tau = 2/3$ photosphere near the top of that atmosphere, which is partially confined from above by free-falling inflowing material (green). The 3 marked radii are the inner boundary of the simulated envelope ($R_0$), the top of the radiative layer ($R_{\rm rad}$), and the photosphere ($R_{\rm \bigstar}$). A thin chromosphere (orange ring) surrounds the photosphere, and its heating-layer photons (gold star and arrows) ionize the wind (teal arrows), where recombination can populate metastable \ion{He}{1}. We also show magnetically confined starspots on the stellar surface, a dense cloud in the extended atmosphere, closed meridional field loops throughout the atmosphere, and a tenuous magnetically heated corona (orange haze) with magnetic reconnection sites (small stars), and $\Omega$ shows the axis of rotation. \label{fig:cartoon}}
\end{figure}

We fit our archival PRISM sources with one uniform configuration with the parameters described above, i.e., a single \texttt{TLUSTY} photosphere plus a host galaxy. Since PRISM cannot resolve the line broadening, we do not broaden the LRD component, and we fix the microturbulence term to 2~km\,s$^{-1}$ and the photospheric metallicity to $[{\rm M/H}] = -1$. The latter is done to ensure that our archival PRISM fits are using the exact same model as our water dots, modulo the additional cold component that we describe below. We then use a free dust screen, which leaves 14 free parameters in total. We fit the 28 DESI sources at each camera's resolution with the metallicity, microturbulence, and broadening left free, giving 18 free parameters for those fits.

Since our 5 primary sources have unique spectral absorption features, we make minor changes to our unified model, most of which are to circumvent library limitations and to account for the fact that two of our primary sources show atmospheric absorption lines. For instance, The Egg and The Rosetta Stone need to have their \ion{Ca}{2} triplets broadened (e.g., \citealt{Juodzbalis2024}; \citealt{Liu2026}) with a line-of-sight velocity dispersion term. We thus fit \textit{The Egg} across all 3 of its arms with a single photosphere, a free microturbulence parameter, a free photospheric metallicity, and an atmospheric broadening term that we confine to $\sim$115--135~km\,s$^{-1}$ (18 free parameters). Since we have $R\gtrsim1000$ spectra for The Egg, we leave the region around its Balmer break unmasked and instead only mask the region around the Ca H and K lines following the recommendations of \citet{Liu2026}. Lastly, we upweight the triplet by $15\times$ to increase the signal coming from those important features, which we do to prevent the continuum pixels from dominating the fit.

We fit \textit{The Rosetta Stone} with the same configuration across its 3 medium-resolution gratings. Since the spectral resolution is much more coarse than the available data for The Egg, we fix the microturbulence term to 2~km\,s$^{-1}$ (the default of the \citet{Liu2026} library). Similarly, since its \ion{Ca}{2} triplet is more tenuous \citep{Juodzbalis2024}, sitting inside the broad Paschen and \ion{O}{1} $\lambda8446$ emission (Section~\ref{sec:obs}), we leave the broadening term free between $50$--$800$~km\,s$^{-1}$ (17 free parameters).

We fit the 3 water dots with the same model that we used to fit the archival PRISM spectra, except that we also include a cold photosphere that shares the same dust screen and adds its own temperature, gravity, and luminosity (17 free parameters each). We fix the metallicity to $[{\rm M/H}] = -1$ due to the fact that it is the only metallicity in our library that goes below $3000$~K, which is approximately the photo-dissociation point of water. Lastly, we upweight the water band $5\times$ for the same reason as the \ion{Ca}{2} triplet above.

\section{Quasi-Stars}
\label{sec:mesaquest}

The innermost layer of the quasi-star is its saturated convection layer. Convection saturates when it carries the largest flux that the gas can transport (i.e., roughly the pressure times the sound speed through each surface) and is thus the limit at which convection can sustain super-Eddington energy transport \citep{Begelman2001} and convection-dominated accretion flows \citep{Quataert2000}. Quasi-stars are thought to occur when the core of a supermassive star formed inside a rapidly inflowing gas cloud collapses, eventually forming a newborn black hole after Myr timescales. Accretion onto this black hole then puts more power into the surrounding gas than ordinary convection can carry. This drives the innermost envelope to saturation, forming a stellar envelope (e.g., \citealt{Begelman2006, Begelman2008, Begelman2010, Ball2011, Ball2012, Coughlin2024}). Notably, the saturated convection layer has an analytical solution \citep{Coughlin2024}, and, more recently, the outer envelopes have been modeled with stellar evolution codes using the analytical solution of the saturated convection as an inner boundary condition (e.g., \citealt{Ball2011, Ball2012, Hassan2025, Santarelli2026}). 

Figure~\ref{fig:cartoon} is a cartoon of our model (not to scale). Quasi-stars can drive outflows off of the radiative layer \citep{Fiacconi2016,Santarelli2026}, and we predict that the material this outflow lifts can form an extended atmosphere whose $\tau = 2/3$ surface is the (pseudo-)photosphere that we observe, much in the same way that stellar envelopes near the Eddington limit extend while staying hydrostatic (see, e.g., \citealt{Grafener2012} for an application of this to luminous blue variables and Wolf-Rayet stars). The in-falling birth-cloud material can confine that atmosphere from above \citep{Begelman2008, Begelman2010}, while also extending the life of the quasi-star by adding mass to the envelope \citep{Santarelli2026, Hassan2025, RomanGarza2026}. We add to this picture several other components of stellar atmospheres, such as starspots, heating layers, clouds, field loops, and a chromosphere, each of which we discuss in the following sections or in Paper~II. In brief, the saturated convection layer is determined by theory, \texttt{MESA} then models the adiabatic and radiative layers out to its own $\tau = 2/3$ surface, and we treat the wind with the mass-loss model of Section~\ref{sec:mass-loss} and the extended atmosphere with the conservation laws of Section~\ref{sec:winds}. We do not claim that every LRD has each of these components; this cartoon is meant to be an illustration that highlights the diversity of phenomena that could be affecting LRD spectra. 

We note that none of the emission or absorption features in Figure~\ref{fig:cartoon} are formed below the photosphere since electron scattering in the atmosphere broadens the lines and a chromosphere can supply the modest ionizing power that observed helium lines need (see also Paper~II). Models that keep an active galactic nucleus at the center of an optically thick cocoon instead require a low-density funnel or a partially covering gap through which the central ultraviolet escapes to a broad-line region above the photosphere \citep{LiuFengHo2026, MadauMaiolino2026}. A quasi-star cannot sustain that geometry since, dynamically, the outermost layers are highly convective and radiation dominated \citep{Coughlin2024,Hassan2025}. The optical depth of the overlying material can reach $\tau \approx 10^{4}$ where the free-fall time is six years at the inner boundary and 25 at the photosphere while convection has mixing timescales on the order of decades, so any hole in the envelope would quickly reseal within years. 

\subsection{The \texttt{MESA-QUEST} Simulations}

We evolve the model described in Section~\ref{sec:mesaquest} with \texttt{MESA-QUEST} \citep{Campbell2025, Santarelli2026, Santarelli2026b}, which is a modification of the \texttt{MESA} stellar evolution code that replaces the innermost boundary of the hydrostatic envelope with conditions that describe an accreting central black hole. The code defines the coordinate of the inner mass as $m(R_0) = M_{\rm BH} + M_{\rm int}$, which includes the gas interior to the inner boundary ($M_{\rm int}$), and the inner luminosity as $l(R_0) = L_{\rm BH}$ at the inner boundary radius $R_0$. The outer boundary is the standard \texttt{MESA} photospheric condition in which the temperature at the \texttt{MESA-QUEST} model's $\tau = 2/3$ surface (i.e., the top of the radiative layer $R_{\rm rad}$) is described by a gray Eddington $T(\tau)$ relation, and the total surface pressure is $(2/3)(g/\kappa)(1 + \Gamma_{\rm ph})$, where $\Gamma_{\rm ph}$ is the local Eddington ratio of the escaping flux. Every model is therefore always in hydrostatic equilibrium by construction. The injected luminosity is
\begin{equation}
  L_{\rm BH} = \alpha_{\rm acc}\,\frac{4\pi G M_{\rm rad} c}{\kappa_0},
  \label{eq:LBH}
\end{equation}
where $c$ the speed of light, $\kappa_0$ is the Rosseland mean opacity of the innermost zone, and $\alpha_{\rm acc}$ is a dimensionless accretion factor \citep{Santarelli2026}.\footnote{We note that \citealt{Santarelli2026} use $\alpha$ for this term, but we use $\alpha_{\rm acc}$ to separate it from the mixing-length theory $\alpha_{\rm mlt}$.} The corresponding BH growth rate is $\dot{M}_{\rm BH} = (1-\varepsilon)L_{\rm BH}/(\varepsilon c^2)$, and we use the radiative efficiency $\varepsilon = 0.1$ throughout.

Our models use \texttt{MESA} r25.12.1 with the low-temperature molecular opacity table AESOPUS \citep{Marigo2009}, which include the H$_2$O, TiO, and CO bands that dominate at our photospheric temperatures of interest. We seed the runs with supermassive stars of $10^4$, $10^5$, and $10^6\,M_\odot$, and distribute their metals at solar abundance ratios using the same metallicities that our \texttt{Prospector} fit returns. We do not seed below this level since a quasi-star envelope must stay radiation-pressure dominated \citep{Begelman2008, Coughlin2024}. Independent testing shows that this assumption breaks down at lower masses, where the ratio between the gas pressure and the total pressure rises to $\beta\sim10\%$ for $10^3\,M_\odot$ seeds and $\beta\sim30\%$ for $10^2\,M_\odot$ seeds, the latter of which often fails to initialize entirely. Similarly, \citet{RomanGarza2026} find that the general-relativistic instability that allows a supermassive star to transition into its quasi-star phase is only possible at envelope masses $\gtrsim10^4M_\odot$, implying that our seed masses align with what can be created by a supermassive star. We evolve our simulations until the black hole mass exceeds $M_{\rm BH}/M_{\rm rad} \approx 0.33$, above which the quasi-star solution breaks down \citep{Coughlin2024, Hassan2025}. That limit is due to the fact that a quasi-star radiates near the Eddington limit of the total star, so the envelope must stay massive enough to confine the outgoing radiation hydrostatically. The exact ratio depends on how the boundary condition is defined; \citet{Coughlin2024} analytically derive limits of $M_{\rm BH}/M_{\rm rad} \approx 0.55$--$0.62$, while the \texttt{MESA} simulations run by \citet{Hassan2025} find this instability occurs around $0.33$, and we adopt the latter since we similarly evolve our quasi-stars with \texttt{MESA}.

\subsection{Mass Loss}
\label{sec:mass-loss}

Quasi-stars drive outflows off of the radiative layer such that the $\tau = 2/3$ surface that we observe instead sits atop an extended atmosphere similar to those seen around asymptotic giant branch (AGB) stars (see below). We use the word wind for the flow that escapes above the photosphere, following the usual meaning of the term for hot stars \citep{LamersCassinelli1999}, and we call the slower material beneath the photosphere the extended atmosphere, after the levitated layers of cool AGB stars \citep{Hofner2018, Gonzalez-Tora2023}. In steady state, the mass that leaves the radiative layer travels through the extended atmosphere and escapes in the wind at the rate $\dot{M}_{\rm wind}$.

\texttt{MESA-QUEST} defines super-Eddington mass loss using a power-law fit to the radiation-driven outflow solutions of \citet{Fiacconi2016}, i.e.,
\begin{equation}
  \dot{M}_{\rm wind} = 1.4\times10^{-4}
    \left(\frac{M_\bigstar}{M_\odot}\right)^{0.96}
    \left(\frac{M_{\rm BH}}{M_\odot}\right)^{0.17}
    M_\odot\,{\rm yr}^{-1}.
  \label{eq:mdot_fr}
\end{equation}
Equation~\eqref{eq:mdot_fr} is an upper-limit on the maximum radiation-pressure driven mass loss rate possible under the assumption that there is no competing in-falling ram pressure that can hold down the outflow \citep{Fiacconi2016, Santarelli2026}. Although this term was originally derived to describe the radiation-driven outflows coming off of the radiative layers of quasi-stars, we use it interchangeably to describe the wind coming off of the extended atmosphere since both regions are dominated by radiation. Modeling such winds as a steady, spherically symmetric flow at its terminal speed $v_\infty$, with density $\rho(r) = \dot{M}_{\rm wind}/(4\pi r^2 v_\infty)$ thus yields 
\begin{equation}
\dot{M}_{\rm wind} \leq \frac{8\pi\,v_\infty R_{\rm \bigstar}}{3\,\kappa_{\rm es}}.
\label{eq:rwind}
\end{equation}

\subsection{The Eddington Ratio and LRD Observables}
\label{sec:phi}

Spectral fitting allows us to constrain effective temperatures and surface gravities, but not masses or radii. To attain the latter, we define two Eddington ratios:
\begin{equation}
  \Gamma_{\rm es} \equiv \frac{L}{L_{\rm Edd}(\kappa_{\rm es})}, 
  \qquad \phi \equiv \frac{\kappa_{\rm es}\sigma T_{\rm eff}^4}{g_{\rm net}c},
\label{eq:phi}
\end{equation}
where $L$ is the bolometric luminosity, $L_{\rm Edd} = 4\pi G M_\bigstar c/\kappa_{\rm es}$ is its Eddington luminosity, $\sigma$ the Stefan--Boltzmann constant, and $g_{\rm net}$ is the fitted gravity of Section~\ref{sec:TLUSTY}. That is, while $\Gamma_{\rm es}$ is a physical property of the photosphere, $\phi$ is what we actually observe due to the effects of $g_{\rm dyn}$ described above. The two are thus equivalent only when $g_{\rm net}\approx g$ via $L = 4\pi R^2 \sigma T_{\rm eff}^4$ and $g = GM/R^2$, which makes $\phi$ a spectroscopic approximation for $\Gamma_{\rm es}$. Any spectroscopic fit therefore places a source on an iso-$\phi$ line in the $\log g$--$T_{\rm eff}$ plane. We call the variable $\phi$ rather than $\Gamma_{\rm es}$ due to this approximation, and we describe below how other factors like rotation or magnetic support can shift $g_{\rm net}$ away from $g$. 

Our \texttt{MESA-QUEST} simulations are super-Eddington throughout most of their envelope in the sense that their local Eddington ratios rise above unity throughout most of the adiabatic and radiative layers, which creates density inversions that fill most of the interior (see Appendix~\ref{app:gamma}). A \texttt{MESA} simulation cannot follow that behavior out to the surface, however, since a hydrostatic code cannot produce a wind (see Appendix~\ref{app:gamma}). What each simulation returns at its photosphere is therefore the surface that a quasi-star would have in the absence of any mass loss. \texttt{MESA-QUEST} therefore only models the adiabatic and radiative layers of Figure~\ref{fig:cartoon}, while our fitted temperatures and gravities measure the extended atmospheres. We discuss in Section~\ref{sec:winds} how our $\phi$ definition can let us measure how much the outflow expands the extended atmosphere. 

\begin{figure*}[t!]
\centering 
\includegraphics[width=0.95\textwidth]{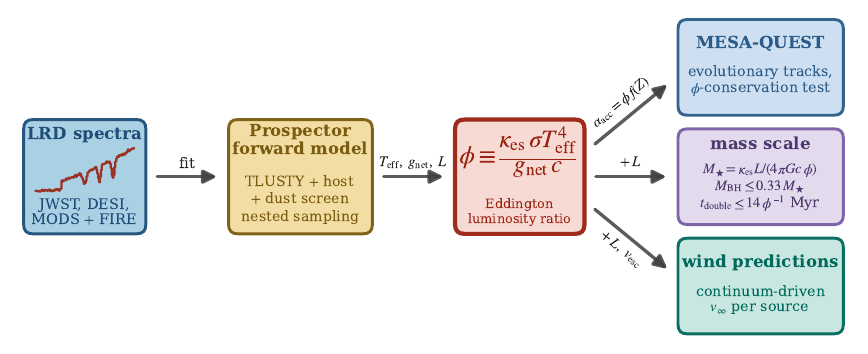}
\caption{Flowchart of the methods presented in this manuscript. We first fit each spectrum with a \texttt{Prospector} model that includes one (or multiple) \texttt{TLUSTY} photospheres plus a host galaxy, which we use to derive the Eddington luminosity ratio $\phi$ from the fitted effective temperatures, surface gravities, and luminosities of our quasi-stars. The rest of this manuscript uses this parameter to test various aspects of the quasi-star hypothesis. For instance, we match $\phi$ to the accretion factor $\alpha_{\rm acc}$, which allows us to more closely simulate each of our sources with \texttt{MESA-QUEST}. We can also use $\phi$ to derive envelope masses, black hole masses, and black hole doubling times, and, lastly, we can derive escape velocities and terminal wind speeds from each of our fitted photospheres. \label{fig:pipeline}}
\end{figure*}

\subsubsection{Mapping Observations to Simulations}
\label{sec:mapping}

A shared premise between the analytical and numerical treatments of quasi-stars is the fact that they radiate at approximately the Eddington limit of the whole envelope \citep{Ball2011, Coughlin2024, Hassan2025, Santarelli2026}. This quantity, $\alpha_{\rm acc}$ in the \texttt{MESA-QUEST} formalism, maps to our fitted $\phi$ values through
\begin{equation}
  \alpha_{\rm acc} = \phi\,f(Z), \qquad f(Z) \equiv \frac{\kappa_0(Z)}{\kappa_{\rm es}},
  \label{eq:alpha_fz}
\end{equation}
where $\kappa_0$ is the Rosseland mean opacity of the innermost cell that is simulated by \texttt{MESA-QUEST} (i.e., the base of the adiabatic convection layer; \citealt{Santarelli2026}), and $f(Z)$ is its ratio to $\kappa_{\rm es}$. A metal-free interior is hot enough that $\kappa_0 \approx \kappa_{\rm es}$ and $\phi \approx \alpha_{\rm acc}$, but metals can raise $\kappa_0$ through bound-free and free-free opacity. We calibrate $f$ on our converged models using the value of $\kappa_0$ that we derive at time $t=0$, finding that $f(Z) \approx 1.15$. However, as the simulations evolve, the opacity of the innermost cell evolves such that any observed $\phi$ only fixes $\alpha_{\rm acc}$ to within $\pm20\%$, which we propagate throughout the rest of our calculations. 

The value of $L_{\rm Edd}$, and with it Equation~\eqref{eq:alpha_fz}, depends on the opacity and the layer (or \texttt{MESA}) cell that we read it from. Written as a function of radius, the local radiative component of the Eddington ratio is
\begin{equation}
  \Gamma_{\rm rad}(r) = \frac{\kappa(r)\,\ell_{\rm rad}(r)}{4\pi G m(r) c},
  \label{eq:gammalocal}
\end{equation}
where $\kappa(r)$ is the local Rosseland mean opacity, $\ell_{\rm rad}(r)$ is the radiative luminosity, and $m(r)$ is the enclosed mass. Quasi-star models evaluate this ratio either at the base of the radiative layer in the hot, metal-poor limit $\kappa_0 \approx \kappa_{\rm es}$ \citep[e.g.,][]{Coughlin2024} or at the base of the adiabatic convection layer, where $\kappa_0$ rises to $0.5$--$0.75$~cm$^2$\,g$^{-1}$ as the envelope evolves \citep{Hassan2025, Santarelli2026}.

Super-Eddington in our usage therefore means that $\phi > 1$ for the quasi-star as a whole, referenced to its total mass $M_\bigstar$. It does not imply a lack of hydrostatic equilibrium, and every source that we fit with \texttt{MESA-QUEST} is indeed modeled in equilibrium due to the nature of the code. The corresponding local radiative ratio at the top of the radiative layer is $\Gamma_{\rm ph}(R_{\rm rad}) \approx \phi\,\kappa(R_{\rm rad})/\kappa_{\rm es}$, which is reduced by the small photospheric Rosseland opacity $\kappa(R_{\rm rad}) \ll \kappa_{\rm es}$.

We use Equation~\eqref{eq:alpha_fz} to map observations to simulations throughout. Appendix~\ref{app:corrections} takes this mapping a step forward, showing how this ratio behaves in hydrostatic equilibrium under the effects of rotation, magnetic-support, and inflow--outflow corrections, where we confirm that none of these assumptions change our final black hole masses by more than a factor of two. Equation~\eqref{eq:alpha_fz} therefore makes $\phi$ a fitted quantity that depends only on the accretion factor and the metallicity, $\phi = \alpha_{\rm acc}/f(Z)$, at every mass and age. Our fitted temperatures and gravities therefore measure $\alpha_{\rm acc}$, and we can use the fitted luminosities to derive masses through Equation~\eqref{eq:mass} and radii through the Stefan--Boltzmann law. In the $T_{\rm eff}$--$\log g$ plane, a quasi-star of known $\alpha_{\rm acc}$ and metallicity must sit on the iso-$\phi$ line
\begin{equation}
    \log g = \log\!\left[\frac{\kappa_{\rm es}\,\sigma T_{\rm eff}^{4}\,f(Z)}{\alpha_{\rm acc}\,c}\right].
    \label{eq:isophi}
\end{equation}

\subsection{$\phi$ Conservation in an Extended Atmosphere}
\label{sec:winds}

Luminosity must be conserved from the top of the radiative layer to the photosphere at the top of the extended atmosphere, at radius $R_{\rm \bigstar}$, such that the Stefan--Boltzmann law and Newtonian gravity relate the effective temperature $T_{\rm eff}$ and Newtonian gravity $g$ at the photosphere to the temperature $T_{\rm eff,rad}$, gravity $g_{\rm rad}$, and radius $R_{\rm rad}$ of the radiative layer as
\begin{equation}
  T_{\rm eff} = T_{\rm eff,rad}\left(\frac{R_{\rm rad}}{R_{\rm \bigstar}}\right)^{1/2},
  \qquad
  g = g_{\rm rad}\left(\frac{R_{\rm rad}}{R_{\rm \bigstar}}\right)^{2}.
  \label{eq:windshift}
\end{equation}
These relations are true only if there are no other sources of luminosity and that very little energy is lost through winds (see Appendix~\ref{app:corrections}). Under hydrostatic equilibrium, $g_{\rm net}\approx g$ (see Section~\ref{sec:TLUSTY}), and substituting Equation~\eqref{eq:windshift} into Equation~\eqref{eq:phi} shows that the factors of $R_{\rm \bigstar}$ cancel, providing a relation between $\phi$ and its corresponding value at the top of the radiative layer, $\phi_\star$, as
\begin{equation}
  \phi = \frac{\kappa_{\rm es} L}{4\pi G M_\bigstar c} = \phi_\star.
  \label{eq:phi_conserved}
\end{equation}
An extended atmosphere, under the assumption that $g_{\rm net}\approx g$, thus conserves the Eddington ratio. The $\tau = 2/3$ surface then moves outward, so the photosphere that we observe is cooler and has a lower gravity than the radiative layer beneath it, much in the same way that the extended atmospheres of AGB stars are cooler than their interior radiative layer. In some cases, we can see both layers at once only if the extended atmosphere has a non-unity covering fraction (see Section~\ref{sec:res_mass}).

Lastly, in Figure~\ref{fig:pipeline}, we show a flowchart of all of the methods presented in this section. Reading from left to right, we begin with a spectrum of an LRD that we feed into our \texttt{Prospector} model, the latter of which includes both a host galaxy and a \texttt{TLUSTY} photosphere. We then use the fitted $(L,T_{\rm eff}, \log g)$ set to derive $\phi$, which we use throughout the rest of the manuscript to run individual quasi-star simulations.

\section{Results}
\label{sec:results}

\subsection{Photospheric Fits and Eddington Ratios}
\label{sec:res_fits}

\begin{deluxetable}{lllcccc}
\tablecaption{Photospheric parameters and Eddington ratios of the fiducial fits. Here, the uncertainties show the spacings to the adjacent TLUSTY grid nodes. The GN-28074 uncertainties are expanded to include its degenerate young-host solution (see Section~\ref{sec:res_pop}). \label{tab:phi}}
\tablehead{
  \colhead{Source} & \colhead{Redshift} & \colhead{Component} & \colhead{$T_{\rm eff}$ (K)} &
  \colhead{$\log g$} & \colhead{[M/H]} & \colhead{$\phi$}
}
\startdata
 J1025$+$1402 (The Egg) & $0.1007$ & single & $4502^{+498}_{-2}$ & $-3.05^{+0.04}_{-0.06}$ & $-1$ & $299^{+40}_{-26}$ \\
    WIDE-EGS-2974 & $2.3203$ & hot & $4379^{+121}_{-379}$ & $-2.57^{+0.21}_{-0.18}$ & $-1$ & $87^{+41}_{-31}$ \\
     & & cold & $2110^{+890}_{-110}$ & $-0.23^{+0.46}_{-0.63}$ & $-1$ & $0.022^{+0.067}_{-0.014}$ \\
    UNCOVER-A2744-20698 & $2.42$ & hot & $3994^{+6}_{-994}$ & $-3.22^{+0.07}_{-0.07}$ & $-1$ & $273^{+49}_{-34}$ \\
     & & cold & ${2015^{+985}_{-15}}$ & ${-1.07^{+0.64}_{-0.66}}$ & $-1$ & ${0.12^{+0.45}_{-0.096}}$ \\
    CAPERS-UDS-23216 & ${2.2957}$ & hot & ${4512^{+88}_{-12}}$ & ${-2.85^{+0.14}_{-0.26}}$ & ${-1}$ & ${186^{+152}_{-48}}$ \\
     & & cold & ${2166^{+834}_{-166}}$ & ${-0.29^{+0.54}_{-0.71}}$ & ${-1}$ & ${0.029^{+0.11}_{-0.02}}$ \\
    GN-28074 (the Rosetta Stone) & $2.26$ & single & $4110^{+890}_{-110}$ & $-2.91^{+0.67}_{-0.09}$ & $-2$ & $148^{+26}_{-115}$ \\
  \enddata
\end{deluxetable}

\begin{figure*}[t!]
\centering
\includegraphics[width=0.75\textwidth]{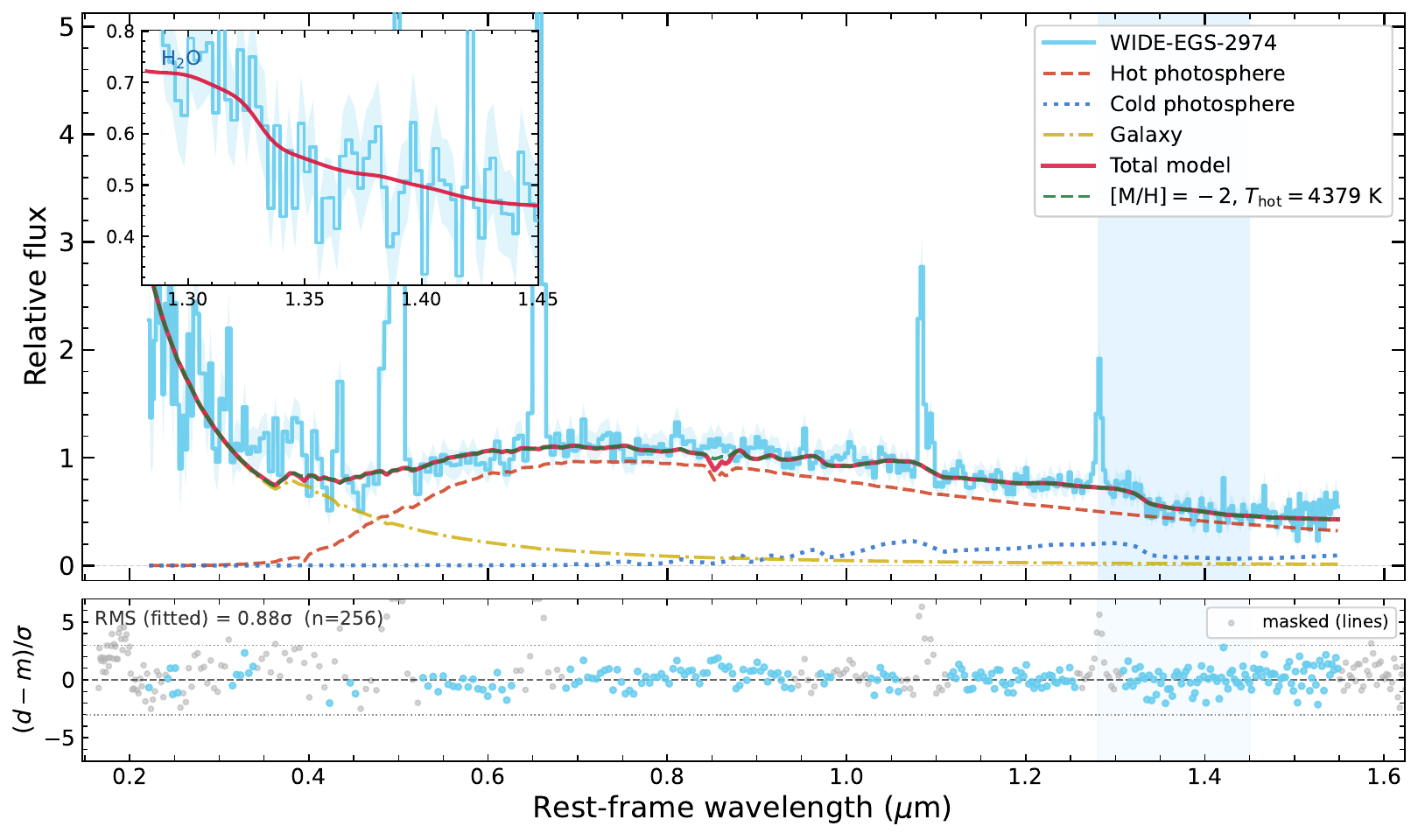}
\caption{Our fiducial two-component fit to WIDE-EGS-2974 ($z = 2.32$). The light blue step curve is the NIRSpec PRISM spectrum with its $1\sigma$ uncertainty band shaded, the red dashed curve is the hot photosphere, the blue dotted curve is the cold photosphere, the gold dash-dotted curve is the host galaxy, and the crimson solid curve is the total model. The two photospheres share a single attenuation. The lower panel shows the fit residuals in units of the uncertainty, with masked regions shaded gray. The light blue vertical band marks the 1.28--1.45~$\mu$m H$_2$O absorption region, which the inset enlarges. We note that the model wants to include \ion{Ca}{2} triplet and Paschen series absorption, producing spurious features on the spectrum. This is purely a modeling constraint that only affects our water dot fits since our \texttt{TLUSTY} models do not include atmospheres with $[{\rm M/H}]<-1$ at the temperatures that water absorption is possible. The dark green curve replaces the hot component with the library's closest metal-poor atmosphere ($[{\rm M/H}] = -2$, which the grid provides only at $4000$~K and above). We also mask the emission lines since our photospheric models are in local thermodynamic equilibrium (LTE) and do not contain emission features. \label{fig:wide}}
\end{figure*}

We measure every $\phi$ in this paper from the fitted temperatures and gravities, with no black hole mass, no bolometric correction, and no assumption about the power source, finding that almost every photosphere that we fit is super-Eddington. Table~\ref{tab:phi} lists our 5 fitted photospheres. With our joint forward model that fits both the host galaxy and quasi-star simultaneously, we provide a solution to the overmassive black hole problem, showing that black holes at the centers of these objects sit at or below local scaling relations, even under the most conservative readings (see Section~\ref{sec:res_mass}). \citet{Begelman2006} and \citet{Begelman2008} predicted objects of exactly this kind, black holes growing inside cool, low-gravity envelopes, nearly a decade and a half before JWST found the first LRD, and our results provide models and simulations that indicate that we may indeed be watching the birth of intermediate-mass black holes (IMBH) with masses between $10^2$--$10^5M_\odot$ \citep{Inayoshi2020}. We show the fit to WIDE-EGS-2974 in Figure~\ref{fig:wide} and the rest in Appendix~\ref{app:spectra}, and we release the chains for each fit derived in this study.\footnote{\url{https://github.com/o-curtis/lrd-quasistar-fits}.}

We fit The Egg with a single photosphere, reproduce its continuum and its atomic absorption lines, and constrain $\phi$ to the $10\%$ level, finding that its photosphere radiates at hundreds of times its own Eddington limit. We also recover a $\phi>100$ super-Eddington photosphere for The Rosetta Stone, which, according to virial mass estimates, historically had one of the largest virial masses in the LRD literature \citep{Juodzbalis2024}. As a quasi-star, it is much like The Egg---a low-mass, late stage, highly eruptive stellar envelope on its way to becoming a typical AGN (see Section~\ref{sec:disc_seq}). However, its spectrum (see Figure~\ref{fig:rs}) still only shows tentative signs of a \ion{Ca}{2} triplet at the current signal to noise, and it has a degenerate, lower $\phi$, lower $M_*$ solution that appears when we assume a different prior on $M_*$ (see Section~\ref{sec:res_pop}), so we treat its fit with the most caution.

\subsection{Water Absorption and the \texttt{MESA-QUEST} Tracks}
\label{sec:res_wind}

The water dots want cold components that are 2--3 dex denser than their hot components, which means that they appear to form in a cold, dense, cloud-like layer rather than at the top of a smooth atmosphere, much like the clouds that can form in the winds of super-Eddington outflows \citep{Shaviv2001} or in compact, pulsation-driven shells of material around a central engine (e.g., \citealt{Cantiello2025, Nandal2026}). We use the word cloud in its gas-phase sense, i.e., as a parcel of dense gas that is confined by the pressure of its surroundings, akin to how the clouds of the interstellar medium or broad-line regions of AGN \citep{Rees1987} are described, and not in the condensate sense as it is usually used in the substellar atmosphere literature (e.g., \citealt{Ackerman2001, Miles2023}) since condensates cannot survive at these temperatures and pressures \citep{Wakeford2017}. We note that starspots are a cool, magnetic alternative to clouds that would also cover part of the surface (Figure~\ref{fig:cartoon}), but, due to the fact that starspots lie on the surface of the star, a 2-component fit to one would return $\phi$ values for each component, so we rule out this interpretation.

\begin{figure}[!htb]
\centering
\includegraphics[width=\columnwidth]{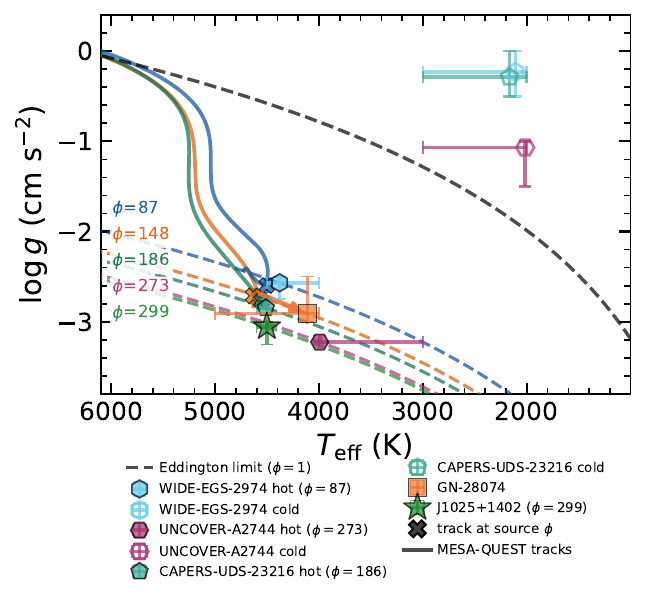}
\caption{The results of our photospheric fits and \texttt{MESA-QUEST} simulations. Here, we plot all fitted components as markers with grid-spacing error bars, the Eddington limit as the heavy dashed curve, and the corresponding iso-$\phi$ lines as light dashed curves. The solid curves are our converged \texttt{MESA-QUEST} evolutionary tracks. Each one enters hot from the upper left and settles onto its stable quasi-star configuration, which we mark by the crosses. The vertical bar on each cross shows a $20\%$ uncertainty in the $f(Z)$ term of Equation~\eqref{eq:alpha_fz}. The arrow  GN-28074 iso-$\phi$ line connects the solution that its simulation settled on at the top of its radiative layer to its fitted photosphere, which, following Section~\ref{sec:winds}, denotes how much the extended atmosphere has been levitated and cooled. The UNCOVER-A2744 and J1025+1402 tracks are not shown since their \text{MESA-QUEST} simulations terminate before their runs reach their corresponding $\phi$ line.}
\label{fig:windshift}
\end{figure}

The cold components being denser than the hot components seemingly imply that the cold components have a higher gas pressure than their surroundings. While magnetic fields could provide this confinement (e.g., \citealt{TakasaoInayoshi2026}), as we discuss briefly in Appendix~\ref{app:corrections}, even modest magnetic fields only change our inferred surface gravities by $\lesssim5\%$. At these low surface gravities, magnetic pressure is possible but would need magnetic fields of order $10^{2}$~G to raise $\log g$ by 2 decades, whereas the equipartition argument that we give in Paper~II shows that these photospheres hold only gauss-level fields, enough to form starspots but not enough to fully confine the gas. We thus conclude that, while starspots cannot be ruled out, complete magnetic confinement is unlikely, so we interpret the cold components as a dense, cold cloud with a non-unity covering fraction formed via ram pressure from shocked wind ejecta (see Section~\ref{sec:disc_popwinds}).

In Figure~\ref{fig:windshift}, we plot all photospheric components in the $T_{\rm eff}$--$\log g$ plane, along with their corresponding \texttt{MESA-QUEST} evolutionary tracks. We use the seed masses of Section~\ref{sec:mesaquest}, and we set the mixing length theory parameter $\alpha_{\rm MLT}$ to 1.0 throughout. These simulations do not, and are not intended to, reproduce the masses and luminosities of these objects. Equation~\eqref{eq:isophi} indicates that the iso-$\phi$ line that each track settles is independent of both, so tuning the models to reproduce $M_\bigstar$ and $L$ precisely would not affect our results.

Each quasi-star starts hot and cools as its black hole grows and its envelope settles, and lands on its own iso-$\phi$ line, where the crosses in the figure denote where it arrives and spends most of its time ($\approx\rm{Myr}$ time scales). The WIDE-EGS-2974 and CAPERS-UDS-23216 tracks settle onto their lines inside the quoted uncertainties of the fitted hot components, while the UNCOVER-A2744-20698 track, like The Egg's, terminates before it reaches its line (see below).

As discussed above, \texttt{MESA-QUEST} does not model how an outflow could build an extended atmosphere, but, in the absence of other sources of luminosity (see Appendix~\ref{app:corrections}), $\phi$ must be conserved through different layers of the atmosphere. As in the case for GN-28074, the path down an iso-$\phi$ line from where the simulation terminates (i.e., at the top of the radiative layer) to the surface that we actually observe and fit is conserved following Equation~\eqref{eq:phi_conserved}. Lastly, we test the robustness of the simulations at high Eddington ratios by raising $\alpha_{\rm acc}$ until the models stop converging, finding that they fail at $\phi \approx 160$--$200$ depending on the mass and mixing length parameter used, which we interpret as a regime that is completely dominated by eruptive mass loss (see Section~\ref{sec:disc_popwinds}) that can no longer be modeled by a hydrostatic code.
 
\begin{figure}[!htb]
\centering
\includegraphics[width=\columnwidth]{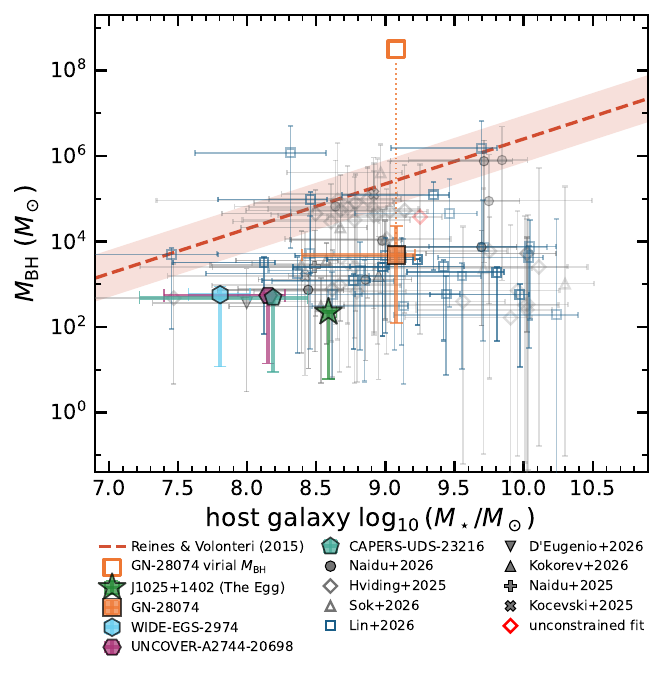}
\caption{Black hole mass against host stellar mass, $M_*$. 75 of the 79 archival sources and all 5 primary sources are consistent with or well below the local scaling relation of \citet{Reines2015}, so no source requires an overmassive black hole. Filled markers show each source's quasi-star mass limit, $M_{\rm BH} = 0.33\,M_\bigstar$, at its fitted host mass, colored green for The Egg, purple for UNCOVER-A2744-20698, light blue for WIDE-EGS-2974, teal for CAPERS-UDS-23216, and orange for GN-28074. The blue pentagons are the DESI fits of \citet{Lin2026}, the best-constrained archival photospheres in our sample 
(Figure~\ref{fig:tlogg}). The gray markers are the archival fits, with one marker shape per catalog. Each marker is the largest mass its black hole can have given the fitted photosphere. The upper error bars show the fitted uncertainties on $M_\bigstar$ while the single bar beneath it spans $M_{\rm BH} = 0.01$ to $0.33\,M_\bigstar$ (Section~\ref{sec:res_mass}). Horizontal bars span the 16th to 84th percentiles of each fitted host mass, including the $\sim0.6$ dex uncertainty that we add in quadrature to account for the effects that our priors have on the fitted $M_*$ values (e.g., Section~\ref{sec:res_pop}). Even with these conservative error bars, every fit sits on or below the \citet{Reines2015} relation---no overmassive black holes are required. Marker opacity scales with uncertainty such that the best-constrained sources appear more opaque. The open square is the virial estimate for GN-28074. The dashed line and band show the local relation of \citet{Reines2015} and its 0.55 dex scatter.}
\label{fig:mass}
\end{figure}
  
\subsection{The Mass Scale}
\label{sec:res_mass}

We now use our fits to calculate the total stellar masses, and, in doing so, individual black hole masses \citep{Coughlin2024, Hassan2025}. For a spherical source, $g = GM_\bigstar/R_{\rm \bigstar}^{2}$ and $L = 4\pi R_{\rm \bigstar}^{2}\sigma T_{\rm eff}^{4}$ such that the stellar mass of the whole quasi-star is then

\begin{equation}
  M_\bigstar = \frac{\kappa_{\rm es}\,L}{4\pi G c\,\phi} = \frac{Lg}{4\pi G\sigma T_{\rm eff}^4},
  \label{eq:mass}
\end{equation}

\noindent where our fits return $M_\bigstar = 700$--$15,000\,M_\odot$ (see also Figure~\ref{fig:mass}). For the water dots, we sum together the fitted luminosities of both their hot and cold components when calculating $L$. A fully opaque cold layer would hide the hot surface entirely, so seeing two components at once requires the cold component to have a non-unity covering fraction, which we can derive by conserving the emitted power per unit solid angle as $f_{\rm cov} \equiv L_{\rm cold}/(L_{\rm hot} + L_{\rm cold}) = 0.16^{+0.04}_{-0.03}$, $0.04\pm0.01$, and $0.08^{+0.03}_{-0.02}$ for WIDE-EGS-2974, UNCOVER-A2744-20698, and CAPERS-UDS-23216, respectively.

We derive black hole masses from the mass limits discussed in Section~\ref{sec:mesaquest}. A quasi-star cannot grow its black hole without limit since the envelope has to keep feeding the inner boundary, meaning the structure fails once the envelope has lost too much material. Bondi-type boundaries stall at a few percent \citep{Ball2012}, saturated-convection interiors can evolve to $0.55$--$0.62$ \citep{Coughlin2024}, and the \texttt{MESA} implementation of \citet{Hassan2025} terminates at $M_{\rm BH}/M_{\rm rad} \approx 0.33$. We adopt the \citet{Hassan2025} relation as the mass limit in Figure~\ref{fig:mass} since it is also derived from \texttt{MESA} simulations, where each filled marker in that figure shows the maximum mass that the black hole could have (i.e., $M_{\rm BH}=0.33M_\bigstar$) with a bar that extends down to $0.01\,M_\bigstar$. Using the saturated-convection ceiling would instead raise the mass limits by $\sim0.27$~dex, while using a Bondi boundary would lower them by nearly a decade, but, even then, neither choice moves the population into the overmassive regime. 
 
\begin{figure}[!t]
\centering
\includegraphics[width=\columnwidth]{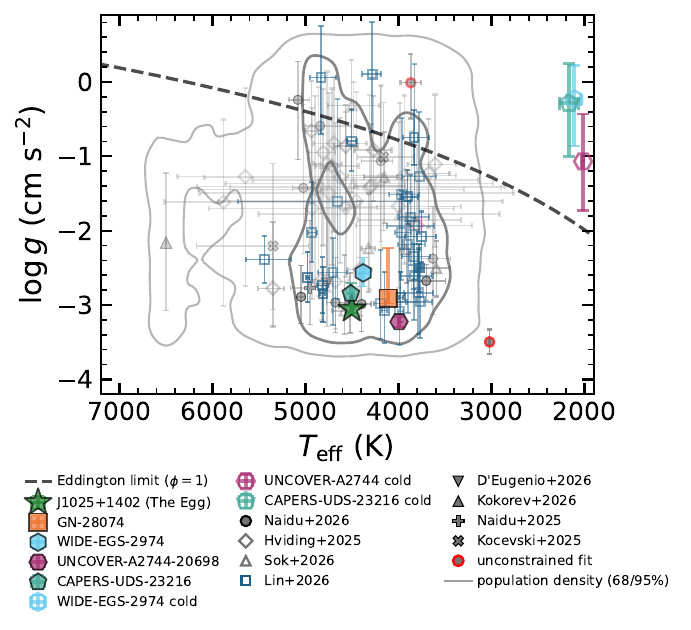}
\caption{Fitted photospheric temperatures and gravities of all LRDs in this work, where the dashed line shows the $\phi = 1$ limit below which every source is super-Eddington. Colored markers are the hot components of our 5 primary sources while the open-faced hexagons show the 3 cold components. The blue pentagons are the DESI fits of \citet{Lin2026}. Gray symbols are the supplementary samples, circles for \citet{Naidu2026}, diamonds for \citet{Hviding2025}, triangles for \citet{Sok2026}, downward triangles for \citet{DEugenio2026}, and upward triangles for \citet{Kokorev2026}. Marker opacity scales with uncertainty such that the best-constrained sources are the most opaque. Error bars span the 16th to 84th percentiles that have been added in quadrature with the 0.32 dex spread in $\phi$ that our host galaxy mass priors impose on the fits (see Section~\ref{sec:res_pop}). The red borders mark the 3 fits whose fits we mark as unconstrained. Contours enclose 68 and 95 percent of the stacked posterior density. \label{fig:tlogg}}
\end{figure}

\subsection{The Population in the Photospheric Plane}
\label{sec:res_pop}

Figure~\ref{fig:tlogg} shows all of our fitted photospheres in the $T_{\rm eff}$--$\log g$ plane. Our 5 primary sources are shown as colored markers, while our supplementary data are drawn in gray. Here, the contours show the inner $68\%$ and $95\%$ spread of the posterior mass, where we have excluded the cold components of the water dots and the unconstrained fits from their derivation. The photospheres concentrate at $T_{\rm eff} \approx 3800$--$5000$~K and $\log g \approx -2.7$ to $-0.8$, and 75 of the 82 median $\phi$ values sit on the super-Eddington side of the $\phi = 1$ line with the population spanning $\phi \approx 0.1$--$370$. The relatively poor spectral resolution of PRISM causes the posteriors of many of these fits to be broad, but our sources still put $\sim82\%$ of their posterior mass in the super-Eddington regime---$5.4\sigma$ higher than the $\sim69\%$ predicted by our priors alone which draw $T_{\rm eff}$ uniformly between $2000$ to $6750$~K and $\log g$ uniformly between $-3.5$ to $+0.5$. Even when we only keep the 28 sources whose $\phi$ posteriors span less than a decade, $\sim91\%$ of the posterior mass has $\phi > 1$, which is $4.4\sigma$ above what our priors predict. Super-Eddington accretion thus seems to be a hallmark of the population that even PRISM-resolution spectra can detect.

We also note that these results are sensitive to the priors that we impose on our host galaxies. On average, our host galaxies are slightly more massive than what \citet{Sun2026} derive by $\sim0.6$ dex. When we refit each galaxy with a host that forms all of its stars within the last 30 Myr, we find no population-level change to our reported $\phi$ values within their quoted uncertainties. However, compared to our fiducial host masses, our fits that impose a low-mass prior on $M_*$ end up producing galaxies that are less massive by $\sim0.65$ dex on average, aligning us with the population average of \citet{Sun2026}. These refits can move individual PRISM $\phi$ values by $\sim0.16$ dex on average, while the DESI values move by $\sim0.32$ dex, so we propagate these shifts into all of our figures, but this choice ultimately changes none of our conclusions. Even with conservative uncertainties, almost all of our fitted LRDs are still consistent with being super-Eddington (Figure~\ref{fig:tlogg}), and almost every quasi-star--host galaxy pair sits well below the \citet{Reines2015} limit (see Figure~\ref{fig:mass}).
  
\section{Discussion}
\label{sec:discussion}

Here, we have applied the quasi-star model to the black hole stars that lie at the center of LRDs \citep{deGraaff2025, Naidu2025, Sun2026}, effectively testing the principles of stellar astrophysics on envelopes with radii $\sim10^{16}$~cm and surface gravities $\log g\lesssim0$. We fit their spectra with a series of \texttt{TLUSTY} stellar atmospheres and model their interiors with the stellar evolution code \texttt{MESA}. In doing so, we derive the relations that convert a fitted temperature and gravity into an accretion rate, a mass, and an escape speed. Similarly to \citet{Gentile2026}, we have made the quasi-star hypothesis testable on individual objects. We show that LRDs are consistent with envelopes akin to those seen in giant stars (similar to the pseudo-photosphere described in \citealt{Ashall2026}). That is, an extended atmosphere with molecular layers, clouds, and, as we will discuss in Paper~II, a chromosphere and a corona (Figure~\ref{fig:cartoon}; \citealt{Hofner2018, Gonzalez-Tora2023, Gonzalez-Tora2024}). All 5 of our primary sources have relatively high $\phi$ values and correspondingly low surface gravities, save for the cold components of the water dots, which likely form in a dense cold cloud that can form in shocked material or super-Eddington-driven radiative-dominated winds (e.g., \citealt{Shaviv2001,Nandal2026}). Molecular and atomic absorption thus seems to be correlated with highly extended LRD atmospheres, though more data are needed to make this claim at a population level.

No source in our sample requires an overmassive black hole, which we show for individual objects (Section~\ref{sec:disc_mass}). We can now start to interpret these results as an evolutionary sequence of quasi-stars (Section~\ref{sec:disc_seq}), and, in doing so, show that our results are consistent with the eruptive mass loss scheme of \citet{Cheng2024} and \citet{Naidu2026}. In our sample, the young, low-$\phi$ envelopes have slow outflows, while the older, least massive envelopes launch outflows at several times their own escape speed (Section~\ref{sec:disc_popwinds}). All of our fitted sources are consistent with the continuum-driven, radiation dominated wind that we describe below in Section~\ref{sec:disc_mass}. With our \texttt{MESA-QUEST} simulations, we generate model quasi-stars that have the same $\phi$ as our fitted photospheres. Our tracks are also among the first quasi-star models evolved at a non-zero metallicity (see also \citealt{RomanGarza2026}), where, compared to pristine simulations, the added metals cool and expand the photosphere (Section~\ref{sec:disc_mass}). Lastly, our fits return the first measured mass function for the population, where envelope mass falls with increasing $\phi$ (Section~\ref{sec:disc_seq}).

\subsection{The Mass Scale in Context}
\label{sec:disc_mass}

For our 5 primary sources, we derive envelope masses between $M_\bigstar\sim10^{2.8}$--$10^{{4.2}}\,M_\odot$ and black holes that must be less than a third of that (Figure~\ref{fig:mass}; \citealt{Hassan2025}), which places these black holes squarely in the intermediate-mass regime that heavy-seed formation channels predict. Appendix~\ref{app:corrections} quantifies how these results depend on some of our underlying assumptions (e.g., the dynamical term $g_{\rm dyn}$, rotation, plane-parallel atmosphere models, and magnetic confinement), where we conclude that our masses are correct to within a factor of a few. Our biggest uncertainty is thus the fact that we only constrain black hole masses to be between $M_{\rm BH}\sim0.01$--$0.33M_\bigstar$ (e.g., \citealt{Coughlin2024, Hassan2025}), so future work is necessary to figure out how to distinguish where along its evolutionary track the quasi-star actually lives.

\begin{figure*}[!t]
\centering
\includegraphics[width=0.9\textwidth]{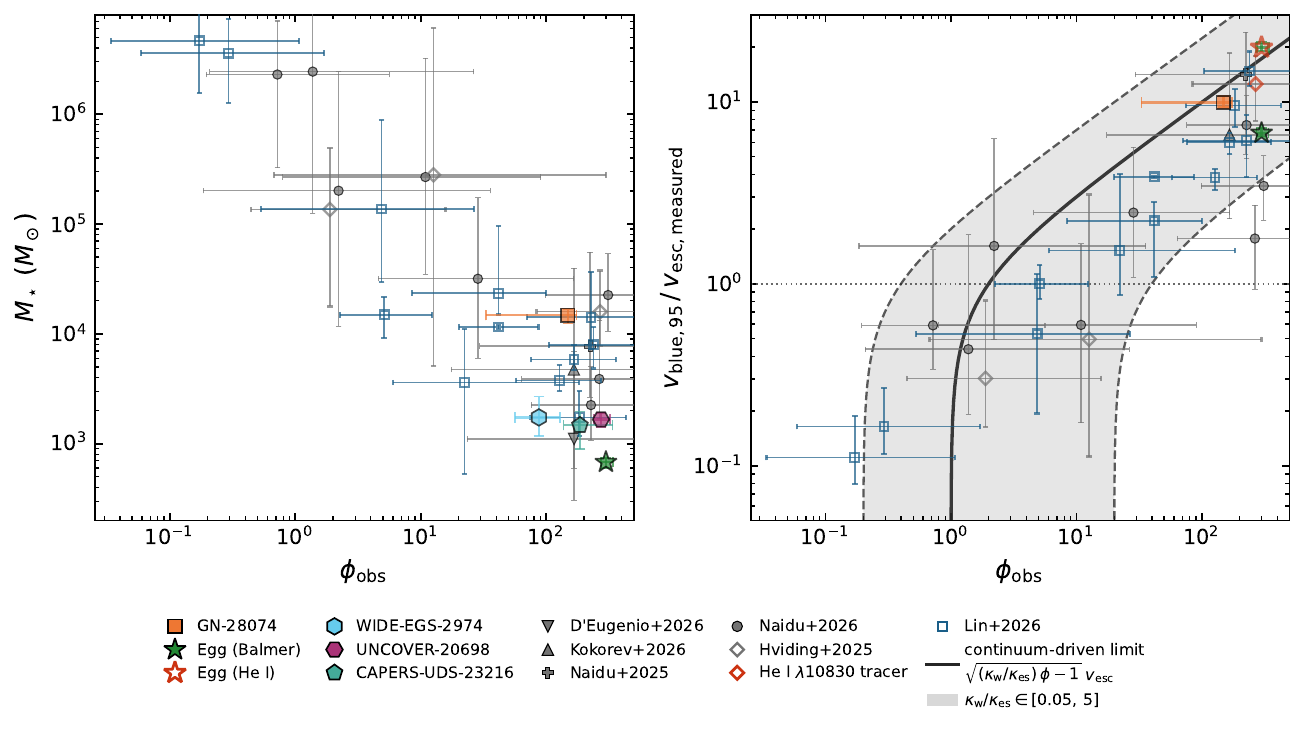}
\caption{\textit{Left:} the envelope mass derived from each source's own posterior through Equation~\eqref{eq:mass}, where the uncertainties show the posterior widths that have been added in quadrature with the $\sim0.16$--$0.32$ dex spread in $\phi$ that our choice of host galaxy stellar mass priors imposes (see Section~\ref{sec:res_pop}). \textit{Right:} $v_{\rm blue,95}$ that we derive from each source's transmission curve (see text and Figure~\ref{fig:pcygni}), normalized by the escape speeds that we derive from each fitted $(L, T_{\rm eff}, \log g)$. When paired with the left panel, we see that the quasi-stars with the lowest $\phi$ values have the most massive envelopes with slow winds relative to their escape speeds, while late-stage quasi-stars have high $\phi$ values and thin, highly eruptive envelopes. The black line shows the continuum-driven limit that radiation dominated, continuum-driven winds can drive as a function of $\phi$ (Equation~\eqref{eq:contdriven}), similar to those seen in Wolf-Rayet stars and classical novae. The gray band varies the ratio between the opacity of the wind and the electron-scattering opacity to values comparable to those seen in the windless photospheres of our \texttt{MESA-QUEST} simulations on the low end and the outflows of line-driven winds on the high end. All of our measured sources are consistent with this radiation dominated, continuum-driven wind model to within $1\sigma$. \label{fig:forecast}}
\end{figure*}

We also resolve the overmassive black hole problem, in which virial estimates place early black holes $10$--$100$ times above the local $M_{\rm BH}$--$M_*$ relation \citep{Harikane2023, Pacucci2023}. For instance, the virial estimate for The Rosetta Stone is $\log(M_{\rm BH}/M_\odot) \approx 8.5$ \citep{Juodzbalis2024}, which is $10^{4.8}$ times higher than the quasi-star mass limits that we place here. \citet{Naidu2026} come to the same conclusion as us, where they discount virial estimates on the grounds that the lines form in a wind through bulk outflow motion and scattering.

Most of the bolometric luminosities in the literature are derived assuming that the line width is generated by the rotational motion around the black hole. For instance, \citet{Juodzbalis2024} convert the broad H$\alpha$ luminosity of The Rosetta Stone into bolometric luminosities using the AGN calibration of \citet{SternLaor2012}, reporting a value that is an order of magnitude above our fitted photosphere ($\sim10^{45}\,\rm{erg}\,s^{-1}$ vs. $\sim10^{44}\,\rm{erg}\,s^{-1}$). \citet{Greene2026} have since measured the full X-ray to far-infrared output of two LRDs, finding that AGN corrections overestimate LRD luminosities by that same order of magnitude since LRDs show neither the X-rays nor the hot dust that the AGN correction factors assume. Using the correction described in \citet{Greene2026}, the H$\alpha$-derived luminosity for The Rosetta Stone agrees with our fit. 
  
The local relation between black hole and host stellar mass \citep{Reines2015}, evaluated at our fitted host masses, predicts black hole masses that are well above our derived masses (Figure~\ref{fig:mass}). In fact, as we show in Section~\ref{sec:res_pop}, even when we refit every LRD with a host that forms all of its stars in a recent ($<30$~Myr) starburst, the population still sits at or below the local scaling relation, and the few sources that rise above it remain consistent with the relation given its intrinsic scatter and the $0.01$--$0.33\,M_\bigstar$ range of each mass limit. Our fits thus return host stellar masses that we do not necessarily believe individually, though they are consistent with the halo masses implied by LRD environment and clustering measurements \citep{Pan2026, CarranzaEscudero2025, Lin2026env} as well as those derived from a host--galaxy decomposition of \citet{Sun2026}. That is to say, for our black holes to become overmassive, we would have to overestimate the host masses of The Egg and The Rosetta Stone, for example, by factors of $\sim300$ and $45$, while even the maximally young host reading of Section~\ref{sec:res_pop} lowers the fitted host masses by only a factor of $\sim$4. We thus conclude that it is difficult to devise a scenario where a quasi-star, or even a black hole star in general, can produce an overmassive black hole. Our archival fits agree with these results, where we find that 75 of the 79 plotted masses fall below the relation, so no source in our sample requires an overmassive black hole, which is exactly the signal that we would expect to see if we truly are observing seeds that are caught partway through their initial growth phases.


Lastly, traditional quasi-star formation is thought to require metal-poor gas, but pristine gas is not necessary to form supermassive stars (e.g., \citealt{ChonOmukai2020}). Most existing models assume primordial gas (e.g., \citealt{Ball2011, Hassan2025, Santarelli2026}) since supermassive star seeds must form in clouds that are too metal poor to fragment, but \citet{ChonOmukai2020} find that such stars can still form at metallicities as high as $\sim10^{-3}\,Z_\odot$, especially when heated under external Lyman-Werner radiation \citep{Bromm2003, Regan2016}. \citet{RomanGarza2026} have even evolved accreting quasi-stars at metallicities up to $Z = 0.01$ finding that their evolutionary tracks change little with $Z$. Several authors (e.g., \citealt{PortegiesZwart2002, Fujii2024, Pacucci2025, Rantala2026}) have shown that runaway stellar collisions in dense star clusters can build supermassive stars of $\sim10^4$--$10^5M_\odot$, which circumvents the traditional need to have quasi-stars form from a primordial gas cloud (e.g., \citealt{Begelman2006}). In fact, if supermassive stars form through such collisions, then quasi-stars should appear wherever young star clusters are forming. This indeed might be the case since LRDs are as common at $z\sim2$ as they are at $z\sim5$ \citep{Kapoor2026, Loiacono2026, Lin2026}, which agrees with the observed number densities of star clusters \citep{Chisholm2026}. We thus evolve our \texttt{MESA-QUEST} simulations at $[{\rm M/H}] = -1$ with the AESOPUS opacities, and, while the added metals cool and expand the photosphere of an evolved quasi-star (Section~\ref{sec:mesaquest}), stable solutions do still evidently exist. 

\subsubsection{Black Hole Doubling Times}

Equations~\eqref{eq:LBH} and \eqref{eq:alpha_fz} connect our inferred Eddington ratios $\phi$ to the black hole accretion factors $\alpha_{\rm acc}$, which we can use to determine how fast the central black hole of the quasi-star is assembling its mass. This is an important thing to consider in the context of supermassive black holes ($\sim 10^{9}\,M_{\odot}$) forming within the first billion years of the Universe's history, i.e., the so-called black hole seeding problem (e.g., \citealt{Inayoshi2020, Regan2024}). Fundamentally, the problem relates to whether ``light seeds'' (i.e., $\sim10$--$100\,M_\odot$ stellar remnants; \citealt{Madau2001}) or ``heavy seeds'' ($\sim10^{4}$--$10^{5}\,M_\odot$ direct-collapse objects; \citealt{Bromm2003}) are responsible for the first supermassive black holes. Light seeds require smaller initial masses but faster doubling times in order to produce the observed supermassive black hole masses at later times, whereas heavy seeds require larger initial masses but fewer doublings (e.g., \citealt{Maiolino2024, Bogdan2024}). Quasi-stars are an example of a heavy-seed channel that grow their black holes at a rate that is proportional to the mass of the entire envelope \citep{Begelman2006, Begelman2008, Hassan2025, Santarelli2026}.

We can estimate the black hole doubling time of our sources using our fitted $\phi$ and $M_\bigstar$ values. Following \citet{Coughlin2024} and \citet{Begelman2026}, we start with the definition of the black hole growth rate from Section~\ref{sec:mesaquest}, i.e., $\dot{M}_{\rm BH} = (1-\varepsilon)L_{\rm BH}/(\varepsilon c^{2})$ for radiative efficiency $\varepsilon$ that we again assume to be $0.1$. We also assume that the energy lost to winds is negligible (see Appendix~\ref{app:corrections}) such that the envelope completely thermalizes the accretion luminosity, giving $L\approx L_{\rm BH}$. If we integrate this relation from time $t=0$ when the black hole has mass $M_{\rm BH}$ to the instantaneous doubling time $t=t_{\rm double}$ when the black hole has mass $2M_{\rm BH}$, we find that $t_{\rm double}=\frac{M_{\rm BH}}{\dot{M}_{\rm BH}}$ \citep{Coughlin2024,Begelman2026}. Equation~\eqref{eq:mass} then gives the total mass of the quasi-star as $M_\bigstar=\frac{\kappa_{\rm es}L}{4 \pi G c \phi}$, which \citet{Hassan2025} relate to the black hole mass by $M_{\rm BH}\leq0.33M_\bigstar$, so we are left with the inequality
\begin{equation}
    t_{\rm double} = \frac{M_{\rm BH}}{\dot{M}_{\rm BH}}
    \leq 0.33\,\frac{\varepsilon}{1-\varepsilon}\,\frac{\kappa_{\rm es}\,c}{4\pi G}\,\frac{1}{\phi}
    \approx 14\,\phi^{-1}~{\rm Myr}.
\label{eq:tdbl}
\end{equation}
The Egg, for instance, has $\phi\sim300$ and will thus double its black hole mass in $\sim50$~kyr, while WIDE-EGS-2974, which has $\phi\sim{90}$, will double in $\sim{0.16}$~Myr. In general, these doubling times range from $\sim$$0.5$--$14$~Myr for our $1\lesssim\phi \lesssim 30$ sources and $\sim$$0.05$--$0.5$~Myr for our $30\lesssim\phi \lesssim 300$ sources. 

\begin{figure}[!t]
\centering
\includegraphics[width=\columnwidth]{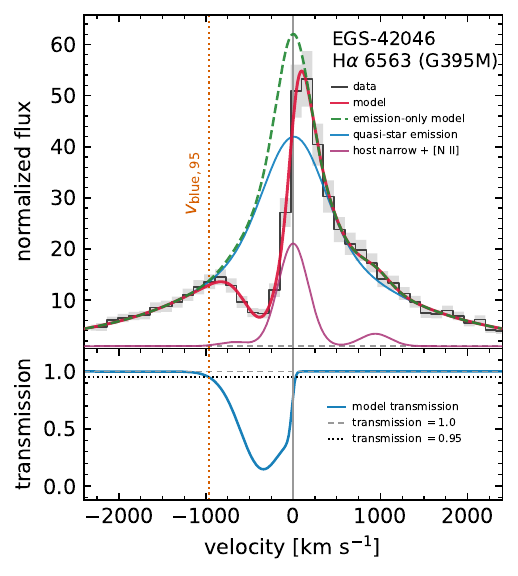}
\caption{Our $v_{\rm blue,95}$ measurement for RUBIES-EGS-42046. The top panel shows RUBIES-EGS-42046's G395M grating spectra of the continuum-normalized H$\alpha$ (solid black) with the best-fit model of \citet{Matthee2026} (solid red; see text), the emission-only model that includes the narrow emission from the host as well as the narrow and broad emission from the quasi-star (dashed green), the quasi-star's emission (solid blue), and the host galaxy's narrow H$\alpha$ and [\ion{N}{2}] emission (solid purple). The bottom panel divides the total model by the emission-only model to produce a transmission curve that represents the total fraction of quasi-star light that has not been absorbed (solid blue). The dotted orange line delineates $v_{\rm blue,95}$, i.e., the blueshifted velocity edge of the absorption trough where the transmission curve returns to 95\% of its systemic value. \label{fig:pcygni}}
\end{figure}

In the absence of outflows, the final masses of these objects are expected to be $\sim$$10^{2}$--$10^{4}\,M_\odot$, approaching the heavy seed scenarios. While winds can strip the envelopes of our highest-$\phi$ sources on the order of $10^{4}$~years (see Appendix~\ref{sec:gdyn}), the presence of inflowing material can resupply their envelopes, offsetting the rate of mass loss (e.g., \citealt{Santarelli2026,RomanGarza2026}). In the context of the evolutionary sequence that we discuss in Section~\ref{sec:disc_seq} below, the young, low-$\phi$, high envelope mass sources double slowly but still have a lot of envelope to either accrete onto their black hole or blow away, while the late-stage, high-$\phi$, lower envelope mass sources are both growing their black holes and expelling their envelopes at much faster rates. 

\subsection{Winds Across the LRD Population}
\label{sec:disc_popwinds}

We now extend our wind analysis beyond our 5 primary sources. From here on, we refer to a source as ``veiled'' when the atmosphere is extended enough that it leaves molecular or atomic absorption (e.g., water absorption) on the spectrum. The Rosetta Stone and The Egg show \ion{Ca}{2} absorption and no molecular bands at $\phi \approx 148$ and $299$, while the water dots show molecular absorption in cold, dense clouds at $\phi \approx 87$--$273$ (Section~\ref{sec:res_wind}). Veiling therefore appears in moderately high-$\phi$ sources whose eruptive winds can expand the atmosphere to low surface gravities while simultaneously compressing part of the outflow into dense, cool clouds. While the majority of our sample shows outflow signatures in their Balmer lines across a range of $\phi$ values, no other source shows strong signs of veiling.

In Figure~\ref{fig:forecast}, we investigate the mass functions and wind signatures across our sample. The left panel shows how our derived envelope masses vary as a function of $\phi$, where we see that the population median is $\sim10^4M_\odot$. These were calculated with Equation~\eqref{eq:mass}, which shows that $M_\bigstar \propto L/\phi$, so the downwards trend is mostly a tautology. Still, we will show below that the least massive sources drive the most eruptive winds, while the most massive sources drive slower outflows.

We first use each of our fitted ($L$, $T_{\rm eff}$, $\log g$) values to calculate the escape velocity of each source as $v_{\rm esc} = \sqrt{2GM_\bigstar/R_\bigstar}$. After subtracting the continuum from our spectra, we follow \citet{Matthee2026} to model the absorbed Balmer profiles of our lines (see Figure~\ref{fig:pcygni} for an example of this decomposition for the H$\alpha$ line of RUBIES-EGS-42046). Using Equation 1 from their paper, we model each of our line profiles as the sum of the emission from the quasi-star that gets partially broadened by electron scattering and absorbed by a skewed-Gaussian absorber, along with the host galaxy's unabsorbed narrow-line component, which, in the case of our region near H$\alpha$, includes the [\ion{N}{2}] $\lambda\lambda6548,6583$ narrow-line doublet. All of this is then convolved with the line-spread function of the instrument. The bottom panel of this figure then shows the ratio of the total model divided by the emission-only model, producing a transmission curve that measures the fraction of light that was not absorbed at a particular velocity. Following \citet{Naidu2026}, we then calculate $v_{\rm blue,95}$, i.e., the blueshifted velocity edge of the absorption trough where the transmission curve returns to 95\% of its systemic value, which we take to be representative of the fastest material that is being launched by each source's wind.

In the right panel of Figure~\ref{fig:forecast}, we plot $v_{\rm blue,95}/v_{\rm esc}$ vs. $\phi$, where a value $>1$ (dashed gray line) indicates that the wind is being launched at a speed greater than its escape velocity. When paired with the left panel, we see that it is the slow, near-systemic absorbers that have the largest masses and the lowest $\phi$ values, while the least massive, highest $\phi$ sources expel extremely eruptive winds. The Rosetta Stone, for instance, has a total mass of $\sim 10^4M_\odot$ and ejects material at nearly $10\times$ its escape velocity. For most of the sources in our sample, the relationship between $v_{\rm blue,95}/v_{\rm esc}$ and $\phi$ very closely follows the continuum-driven limit that radiation dominated super-Eddington winds can drive \citep{Shaviv2001, Owocki2004, vanMarle2008, Quataert2016}. We show this relation in the right panel, where we assume $v_\infty\approx v_{\rm blue,95}$ under the assumption that $v_{\rm blue,95}$ traces the fastest outflowing material of the wind (see also \citealt{Naidu2026}). To derive this relation, consider the fact that the conservation of momentum for a radially symmetric wind dictates that
\begin{equation}
    v\frac{dv}{dr} = -\frac{GM}{r^2} + \frac{\kappa_{\rm w} L}{4\pi r^2 c},
\end{equation}
where the first and second terms on the right hand side are the gravitational acceleration and the outwards radiative acceleration for a wind with opacity $\kappa_{\rm w}$, respectively. Integrating this from positions $r=R_\bigstar$ where the wind has velocity $v=0$ to $r=\infty$ where the wind has velocity $v_{\infty}$ and substituting in the definitions for $v_{\rm esc}^2=\frac{2GM_\bigstar}{R_\bigstar}$ and $\phi=\frac{\kappa_{\rm es}L}{4 \pi G M_\bigstar c}$ gives
\begin{equation}
    \label{eq:contdriven}
    \frac{v_\infty}{v_{\rm esc}} = \sqrt{\frac{\kappa_{\rm w}}{\kappa_{\rm es}}\phi - 1},
\end{equation}
which others have derived variants of in the context of super-Eddington winds in Wolf-Rayet stars and classical novae (e.g., \citealt{Shaviv2001, Owocki2004, vanMarle2008, Quataert2016}). The gray band in Figure~\ref{fig:forecast} varies the $\kappa_{\rm w}/\kappa_{\rm es}$ opacity ratio between $0.05$--$5$. The lower bound is comparable to the opacities that our windless \texttt{MESA-QUEST} photospheres have, while the upper bound represents winds with opacities $>\kappa_{\rm es}$, which is common in line-driven winds \citep{Castor1975}. Every source is consistent with this radiation-dominated, continuum-driven framework to within $1\sigma$.
  
\begin{table*}[!htb]
  \centering
  \caption{Observable predictions of the quasi-star framework from our fitted photosphere parameters, the measurements that test them, and where each test stands as of this work. \label{tab:predictions}}
  \footnotesize
  \begin{tabular}{llll}
  \hline\hline
  \parbox[t]{0.16\textwidth}{Observable} & \parbox[t]{0.33\textwidth}{quasi-star expectation} & \parbox[t]{0.19\textwidth}{Test and requirement} & \parbox[t]{0.22\textwidth}{Status}
  \\[5pt]
  \hline
  \parbox[t]{0.16\textwidth}{Mass scale} & \parbox[t]{0.33\textwidth}{Envelope masses can be derived with $M_\bigstar = \kappa_{\rm es}L/(4\pi G c\,\phi)$ and black hole masses with $M_{\rm BH} \leq 0.33\,M_\bigstar$ (Equation~\eqref{eq:mass}). No overmassive black holes are required} & \parbox[t]{0.19\textwidth}{Population-level photospheric fits with better constrained host masses} & \parbox[t]{0.22\textwidth}{$\phi$ and $M_\bigstar$ measured for {87} individual sources (Sections~\ref{sec:res_pop} and \ref{sec:res_mass})} \\[5pt]
  \parbox[t]{0.16\textwidth}{Wind velocities} & \parbox[t]{0.33\textwidth}{{Winds are radiation dominated and continuum-driven, following the relation $v_\infty/v_{\rm esc} = \sqrt{(\kappa_{\rm w}/\kappa_{\rm es})\phi - 1}$, which rises with $\phi$} (Figure~\ref{fig:forecast})} & \parbox[t]{0.19\textwidth}{Photospheric fits and mid- to high-resolution grating measurements} & \parbox[t]{0.22\textwidth}{Measured for 27 sources, all of which are consistent with an opacity ratio between $\kappa_{\rm w}/\kappa_{\rm es} = 0.05$--$5$ to within $1\sigma$ (Section~\ref{sec:disc_popwinds})} \\[5pt]
  \parbox[t]{0.16\textwidth}{Super-Eddington accretion rates} & \parbox[t]{0.33\textwidth}{Quasi-stars are super-Eddington, so they drive eruptive mass loss that expands their upper atmospheres into the AGB-star-like extended, low surface gravity regime} & \parbox[t]{0.19\textwidth}{Joint host-galaxy + low surface gravity stellar atmosphere fits to measure $\phi>1$ for most sources} & \parbox[t]{0.22\textwidth}{{Confirmed for most of our sample. Even with conservative error estimates, a majority of our sample puts most of their posterior mass into the super-Eddington regime.}} \\[5pt]
  \parbox[t]{0.16\textwidth}{Veiling} & \parbox[t]{0.33\textwidth}{Atomic and molecular absorption occurs in the atmospheres of moderately high-$\phi$, low surface gravity sources. In some cases, cold, dense clouds can cover $\sim10\%$ of their photospheres where water absorption can occur (Section~\ref{sec:disc_popwinds})} & \parbox[t]{0.19\textwidth}{Photospheric fits across the entire population, especially 2-component fits when the spectra show a separate cold component} & \parbox[t]{0.22\textwidth}{Seen in 3 water dots, The Egg, and the Rosetta Stone. The cold components of the water dots are $\sim2$--$3$ dex denser than their fitted hot components} \\[5pt]
  \hline\hline
  \end{tabular}
\end{table*}

Our archival fits also reveal gas that is redshifted rather than blueshifted, which aligns with our interpretation in Figure~\ref{fig:cartoon} which shows the photosphere as the interface between a free-falling inflow and an extended atmosphere akin to those seen in AGB stars (see also Figure~1 of \citealt{Ashall2026} who devises a similar geometry to ours). The Egg, for instance, has an H$\beta$ absorber that is redshifted to $\approx100$~km\,s$^{-1}$ along with its other blueshifted wind lines, which we interpret as the settling of some of the material as it falls back onto the photosphere. 

\subsection{An Evolutionary Sequence for the LRD Population}
\label{sec:disc_seq}

Our measurements suggest an evolutionary sequence within the LRD population. Figure~\ref{fig:forecast} shows that the most massive envelopes have the lowest Eddington ratios and the slowest winds, while the least massive envelopes have the highest Eddington ratios and launch winds at several times their own escape speed. Across the 27 sources with measured absorbers, envelope mass falls with increasing $\phi$. Equation~\eqref{eq:mass} makes $M_\bigstar \propto L/\phi$, so part of this trend is the definition itself, but further testing has revealed that the source luminosities do not trend with $\phi$ (Section~\ref{sec:disc_popwinds}), so the decline reflects in part a real spread in $\phi$ at fixed luminosity rather than the definition alone. We interpret this as an age sequence. A quasi-star begins with most of its birth envelope still in place, so the envelope is massive, less luminous, and its outflows are continuum-driven \citep{Santarelli2026,Santarelli2026b}. Winds progressively strip the envelope while the luminosity remains constant, so $\phi$ climbs and the winds become eruptive \citep{Cheng2024, Naidu2026}. The Egg, for instance, is one of the least massive sources in our sample at, but its envelope is extremely super-Eddington at $\phi = 299$, which places it squarely in the eruptive outflow regime of Figure~\ref{fig:forecast}. 
  
One of the successes of this framework is that it does not require holes or funnels that extend down to traditional broad-line regions in the vicinity of the black hole (e.g., \citealt{LiuFengHo2026, MadauMaiolino2026, Ji2026holes, Tang2026}). Such channels would imprint a viewing-angle dependence on the population, but such dependencies must be minor, since LRDs as a population seem mostly homogeneous \citep{deGraaff2025}. These holes are also unstable in the envelopes that we simulate, where the free-fall and convective timescales are only decades long. A sustained wind or a strong magnetic field could hold a channel open, but, as we discuss in Paper~II, an equipartition argument can reveal that these photospheres can only sustain Gauss level fields at most. The covering fractions that our fits measure are therefore gaps in the extended atmosphere that expose the radiative layer below.

Still, the quasi-star phase must end to give way to an AGN and quasar population, which means that, at some point, sightlines to the interior should open up. \citet{Hassan2025} predict that the quasi-star envelope destabilizes at $M_{\rm BH}/M_{\rm rad} \approx 0.33$, which takes $\sim1$--$20$~Myr depending on mass-loss rate and accretion rate \citep{Santarelli2026b, Hassan2025}. \citet{Begelman2026} described this transitional period in late-stage quasi-stars, and several authors (e.g., \citealt{Pacucci2024, Liu2025, Lambrides2026}) have characterized the final stages as a super-Eddington accretion flow, i.e., a geometrically thick disk or quasi-spherical inflow whose surface radiates at the local Eddington limit. As the mass loss continues, the continuum will start to fade and X-rays and high-ionization lines will start to travel through the atmosphere as the covering fraction falls. As mass loss continues and winds start to become more and more eruptive (Section~\ref{sec:disc_popwinds}), the quasi-star may start to shed its outer envelopes. Much like how late-stage AGB stars can turn into planetary nebulae, quasi-stars may enter a nebular-like phase resembling the gaseous cocoons described by \citet{Sneppen2026cocoon}, which produces strong emission-line sources without the need for the energetic eruptions, pulsations, or convective cells that we describe in Paper~II (see also \citealt{Nandal2026, Naidu2026}). \citet{Hviding2026xrd} may have detected an LRD that has progressed even further into this transition with the detection of X-rays emanating from an LRD-like object. Similarly, the two ``forges'' of \citet{Fu2025} show both the characteristic v-shape spectrum and the X-ray, radio, and hot-dust emission of quasars. The almost negligible X-ray detection rate, which requires Compton-thick columns even for intrinsically X-ray-weak engines \citep{Sneppen2026xray}, implies that this transitory state is rare, even as the LRD number density itself stays roughly flat from $z \sim 5$ to $z \sim 2$ before declining toward lower redshift \citep{Kapoor2026, Loiacono2026}. 

\section{Conclusion}
\label{sec:conclusion}

We fit low-gravity TLUSTY photospheres with \texttt{Prospector} host galaxies to the 5 LRDs with the strongest known absorption features, at $z = 0.1$--$2.4$, as well as to 82 archival JWST and DESI sources. For each, we calculate the Eddington luminosity ratio $\phi = \kappa_{\rm es}\sigma T_{\rm eff}^4/(g_{\rm net}c)$, which we use to derive accretion rate factors, black hole and envelope masses, and escape velocities. Lastly, we interpret each fitted photosphere as a quasi-star, which we simulate using \texttt{MESA-QUEST}. A summary of our results includes:

\begin{enumerate}[leftmargin=*, itemsep=1pt]
\item Most of our sources exhibit super-Eddington luminosity ratios, with our 5 primary sources spanning $\phi={87}$--$299$. All of these correspond to masses between $700$--${15{,}000}\,M_\odot$. Along with the fitted host galaxy stellar masses, we show that no LRD requires an overmassive black hole; we conclude that we are instead observing supermassive black hole seeds. 
\item In the quasi-star scenario, we map the fitted Eddington luminosity ratios to the accretion factor $\alpha_{\rm acc}$ that we use to run \texttt{MESA-QUEST} simulations of each object out to its radiative layer. We have thus made a self-consistent framework that allows us to test the quasi-star hypothesis as a potential central engine for LRDs, where we provide a single, falsifiable model and show that its continuum and absorption features are fully consistent with the energy budget that is transported through quasi-stars.
\item Similarly, we connect $\phi$ to the black hole accretion rates, finding that the doubling time of the black hole at the center of a quasi-star is roughly $t_{\rm double}\leq14\,\phi^{-1}~\rm{Myr}$. Our results indicate that the black holes in the centers of quasi-stars double on timescales of $\sim$$0.05$--$3.5$~Myr and can satisfy the heavy-seed scenario and produce IMBHs within the quasi-star's lifetime, especially in the presence of inflows that resupply their envelopes (e.g., \citealt{Santarelli2026,RomanGarza2026}). 
\item We derive escape velocities for all of our objects, which we compare against measured blueshifted absorption features. Our high-$\phi$, low-mass sources eject material at several times their escape velocities while low-$\phi$, high-mass sources launch slow winds. We interpret this as an evolutionary sequence for LRDs. Recently formed quasi-stars with massive envelopes have low Eddington luminosity ratios, slow winds, and long black hole doubling times, while late-stage quasi-stars have thin, eruptive envelopes and faster black hole doubling times.
\item 3 of our sources show water absorption in their spectra, which necessitates a two-component fit to their data. In all 3, the cold component is colder and denser than its hot component, which we interpret as water forming in dense clouds that cover roughly $4$--$16\%$ of the observed photosphere.
\end{enumerate}

JWST first discovered LRDs within its first year of operations, yet all aspects of their nature, from their broad emission lines, their v-shaped UV-optical spectrum, and their lack of X-ray emission, remain challenging for any single model to describe. And yet, \citet{Begelman2006} and \citet{Begelman2008} first proposed the quasi-star as a possible solution to the supermassive black hole seeding problem nearly two decades before LRDs were widely discovered. Here, we have shown that the defining characteristics of LRDs can consistently be described as originating from the photospheres and extended atmospheres that surround quasi-stars. No LRD in our sample hosts an overmassive black hole, and we have shown that LRDs may provide evidence that heavy seeds, rather than light stellar-remnant seeds, may have produced the supermassive black holes and quasars that we see within the first billion years of the Universe's history (e.g., \citealt{Maiolino2024}). That is, we may be directly observing supermassive black hole seeds. Table~\ref{tab:predictions} summarizes all of the tests that we performed throughout this manuscript. A natural extension of this work is a non-hydrostatic equilibrium numerical simulation, such as a 3D radiation magnetohydrodynamic simulation of a quasi-star's atmosphere and extended envelope that is capable of resolving the convective flux transport, line profiles, eruptive mass launching, and clouds that a hydrostatic solver like \texttt{MESA} cannot model.

\section*{Acknowledgments}
O.C. acknowledges support as a Penn State Extraterrestrial Intelligence Center Postdoctoral Fellow, which is funded through a private donation made by The Ultraintelligence Foundation. O.C. would like to thank Rohan Naidu whose comments helped strengthen this manuscript. O.C. would like to thank Xiaojing Lin for sharing the reduced LBT/MODS and Magellan/FIRE data of The Egg that we used in this analysis, as well as Albert Sneppen for constructive feedback.

Portions of the code used in this work were developed with Claude Code, a subscription to which was paid for by Penn State's Institute for Gravitation and the Cosmos. All of the analysis, text, and claims herein are strictly our own. This research was partially supported by the Seed Grant award ICDS\_READ26\_m1jtw13 from Penn State's Institute for Computational and Data Sciences. The Penn State Extraterrestrial Intelligence Center and the Center for Exoplanets and Habitable Worlds are supported by Penn State and its Eberly College of Science. This research has made use of NASA's Astrophysics Data System Bibliographic Services. We performed all computational work on the Roar Collab supercomputer, administered by Penn State's Institute for Computational and Data Sciences.

J.M.H. acknowledges support from the Evolving Universe Fellowship, which is made possible by a generous donation from Dr. Keiko Miwa Ross. J.M.H. also acknowledges support from JWST Program \#8544.

Some of the data products presented herein were retrieved from the Dawn JWST Archive (DJA). DJA is an initiative of the Cosmic Dawn Center (DAWN), which is funded by the Danish National Research Foundation under grant DNRF140. We retrieved the JWST observations from the Mikulski Archive for Space Telescopes, where the WIDE survey is archived at \dataset[10.17909/57km-j134]{https://doi.org/10.17909/57km-j134}, UNCOVER at \dataset[10.17909/zn4s-0243]{https://doi.org/10.17909/zn4s-0243}, CAPERS at \dataset[10.17909/0q3p-sp24]{https://doi.org/10.17909/0q3p-sp24}, and JADES at \dataset[10.17909/8tdj-8n28]{https://doi.org/10.17909/8tdj-8n28}.
  
\facilities{JWST(NIRSpec), LBT(MODS), Magellan:Baade(FIRE), Mayall(DESI)}

\software{\texttt{MESA} \citep{Paxton2011, Jermyn2023},
\texttt{MESA-QUEST} \citep{Campbell2025, Santarelli2026},
\texttt{TLUSTY} \citep{Hubeny1988, Hubeny2021},
\texttt{Prospector} \citep{Johnson2021},
\texttt{FSPS} \citep{Conroy2009},
\texttt{dynesty} \citep{Speagle2020},
\texttt{AESOPUS} \citep{Marigo2009},
\texttt{astropy}, \texttt{numpy}, \texttt{scipy}, \texttt{matplotlib}, \texttt{sedpy},
Claude Code (Anthropic).}

\appendix
\section{The Local Eddington Structure of the Matched Models}
\label{app:gamma}
\renewcommand{\thefigure}{A\arabic{figure}}
\setcounter{figure}{0}

In this appendix, we measure the local Eddington ratio $\Gamma_{\rm rad}(r)$ of Equation~\eqref{eq:gammalocal} using our \texttt{MESA-QUEST} models to show that the radiative layers of our simulated quasi-stars are indeed sub-Eddington. As a reminder, \texttt{MESA-QUEST} is a hydrostatic code, so all of our simulated objects are sub-Eddington by definition. Still, there are regions in the envelope that have $\Gamma_{\rm rad} > 1$, which creates density inversions that ultimately drive the eruptive mass loss (see, e.g., \citealt{Cheng2024, Hassan2025, Begelman2026} for more detail on eruptive mass loss schemes). \texttt{MESA-QUEST} is only able to model out to the radiative layer of Figure~\ref{fig:cartoon}, so ``photosphere'' in this Appendix refers to this simulated photosphere (whose radius is $R_{\rm rad}$), not to the top of the extended atmosphere that we describe in the main text (whose radius is $R_{\bigstar}$). We show in Section~\ref{sec:winds} that $\phi$ is conserved between these two surfaces, so below, we use our fitted $\phi$ values from the main text.

\begin{figure}[!htb]
\centering
\includegraphics[width=0.4\textwidth]{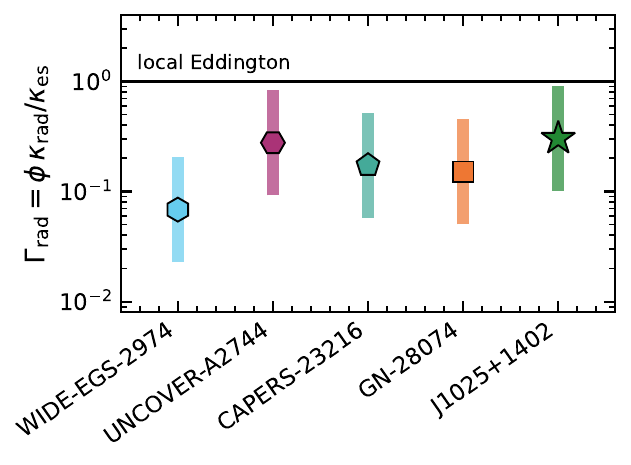}
\caption{Local Eddington ratio at our simulated photospheres, with $\kappa_{\rm rad}$ derived from our \texttt{MESA-QUEST} simulations and $\phi$ from our spectroscopic fits. The horizontal line marks the local Eddington limit. Every simulated photosphere sits below its local Eddington limit, implying that the outflows launch below the photosphere (Section~\ref{sec:phi}).}
\label{fig:gammaladder}
\end{figure}

We evaluate Equation~\eqref{eq:gammalocal} at the photosphere. There, $\Gamma_{\rm rad} = \phi\,\kappa_{\rm rad}/\kappa_{\rm es}$ where $\kappa_{\rm rad}$ is the photospheric Rosseland mean opacity that we calculate from our simulations and $\kappa_{\rm es}$ is the electron-scattering opacity. Figure~\ref{fig:gammaladder} shows the resulting ratios. Each of our simulated photospheres sits below its local Eddington limit.
  
\begin{figure*}[!htb]
\centering
\includegraphics[width=0.7\textwidth]{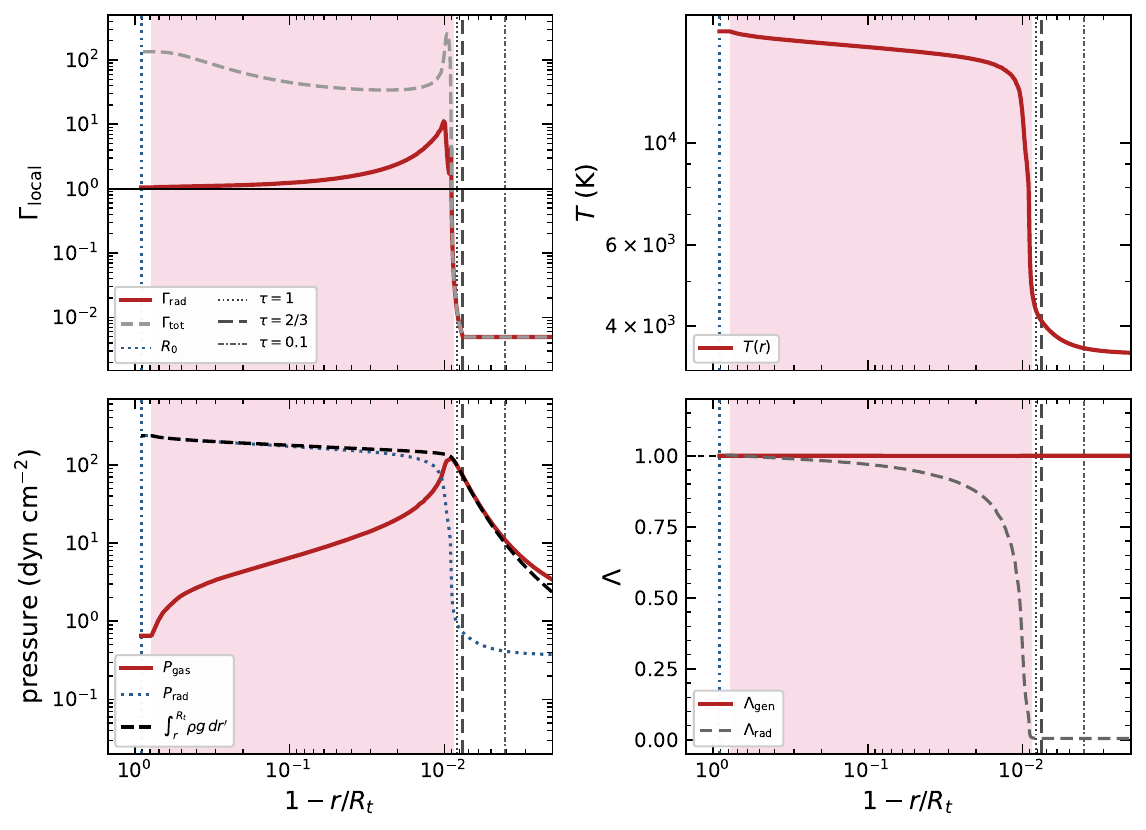}
\caption{Radial structure of the converged $\alpha_{\rm acc} = 13.5$ model at $M_{\rm BH}/M_{\rm rad} = 0.10$ (where $M_{\rm rad}$ is the total mass interior to the radiative layer) and $T_{\rm eff} = 4109$~K. The models end at the top of the radiative layer and we attach the gray hydrostatic atmosphere analytically above it such that $R_t$ is the top of that atmosphere's $\tau = 0.01$ surface. The pink band denotes the density inversions that occur when $\Gamma_{\rm rad} > 1$, and the vertical lines, identified in the top left legend, mark the inner boundary $R_0$ and the $\tau = 1$, $\tau = 2/3$, and $\tau = 0.1$ surfaces. \textit{Top left:} the local Eddington ratios of Equation~\eqref{eq:gammalocal} for the radiative flux (solid) and the total flux (dashed). \textit{Top right:} the temperature profile. \textit{Bottom left:} the gas pressure, the radiation pressure $P_{\rm rad} = aT^4/3$, and the weight of the atmosphere above that layer. \textit{Bottom right:} the launch number $\Lambda$ of Equation~\eqref{eq:launchgen} (solid), which \texttt{MESA}'s hydrostatic assumption forces to be unity, and the radiative share $\Lambda_{\rm rad}$ (dashed).}
\label{fig:gammastruct}
\end{figure*}

We show the internal structure of a \texttt{MESA-QUEST} simulation in Figure~\ref{fig:gammastruct} through a converged $\alpha_{\rm acc} = 13.5$ model at $M_{\rm BH}/M_{\rm rad} = 0.10$ (where $M_{\rm rad}$ is the total mass interior to the radiative layer) and $T_{\rm eff} = 4109$~K. Deep in the envelope, the radiative flux sits at the local Eddington value. The shaded band shows where $\Gamma_{\rm rad}$ exceeds unity, which leads to a density inversion. Outside the band, the opacity tends towards $\kappa_{\rm rad} \approx 10^{-4}\,\rm{cm}^{2}\,g^{-1}$ and the model's surface sits at $\Gamma_{\rm rad} \approx 5\times10^{-3}$, locally sub-Eddington. The temperature panel shows the same structure, where the $\tau = 1$ layer is $\sim6\%$ hotter than the $\tau = 2/3$ surface, implying limb-darkening across the photosphere.

We then derive a launch criterion for the atmosphere. That is, we derive the condition for which any outgoing pressure cannot be contained by the weight of the atmosphere above it. To do this, we first compute the gas pressure $P_{\rm gas}$ and the radiation pressure $P_{\rm rad} = aT^4/3$. By the conservation of momentum, the gas in a spherically symmetric envelope must obey
\begin{equation}
  \rho\,\frac{Dv}{Dt}
    = -\frac{\partial}{\partial r}\left(P_{\rm gas} + P_B\right)
      - \frac{Gm\rho}{r^2}
      + \frac{\kappa\rho\,\ell_{\rm rad}}{4\pi r^2 c}
      + \rho\,\Omega^2 r,
  \label{eq:momentum}
\end{equation}
where $v$ is the radial velocity, $P_{\rm gas} = \rho k_B T/(\mu m_p)$ is the gas pressure for temperature $T$, Boltzmann constant $k_B$, mean molecular weight $\mu$, and proton mass $m_p$, $P_B = B^2/8\pi$ is the magnetic pressure for field strength $B$, $\Omega$ is the rotation rate, and the third term is the radiative force per unit volume, $\Gamma_{\rm rad}\,\rho g$.

We then integrate Equation~\eqref{eq:momentum} from $r$ to the top of the gray atmosphere ($R_t$) and define $W(r) = \int_r^{R_t}\rho g\,dr'$ as the total weight of the atmosphere above the layer $r$, giving the launch criterion $\Lambda(r)$ as
\begin{equation}
  \Lambda_{}(r) = \frac{1}{W(r)}\left[\int_r^{R_t}\Gamma_{\rm rad}\,\rho g\,dr' + \Delta P_{\rm gas} + \Delta P_B + \int_r^{R_t}\rho\,\Omega^2 r'\,dr'\right],
  \label{eq:launchgen}
\end{equation}
where $\Delta P_{\rm gas} = P_{\rm gas}(r) - P_{\rm gas}(R_t)$ and $\Delta P_B$ is its magnetic counterpart. Launching a wind requires $\Lambda > 1$. Although written generally here, the magnetic and rotational terms are negligible and set to 0 throughout our calculations.

Since \texttt{MESA} is a hydrostatic code, $\Lambda(r)$ is trivially 1 throughout the entire envelope to within $3\times10^{-4}$. However, we can now separate out the radiative and gas components. The radiative component, $\Lambda_{\rm rad}(r) \equiv W^{-1}\!\int_r^{R_t}\Gamma_{\rm rad}\,\rho g\,dr'$, is the fraction of the atmosphere's weight above $r$ that is supported by the radiation. If the rotational and magnetic terms are negligible, then, in equilibrium, the launch condition is satisfied when $\Delta P_{\rm gas} = W\,(1 - \Lambda_{\rm rad})$.

\section{Higher-Order Corrections to the Mass Scale}
\label{app:corrections}
\renewcommand{\thefigure}{B\arabic{figure}}
\setcounter{figure}{0}
\renewcommand{\thetable}{B\arabic{table}}
\setcounter{table}{0}

This appendix describes further corrections that one could apply to the $\alpha_{\rm acc} = \phi\,f(Z)$ relation, all of which can affect the black hole masses that we derive. In brief, we quantify how $g_{\rm dyn}$, magnetic support, a plane-parallel bias, and rotation affect our observed $\phi$ values, and we conclude that none of these phenomena changes our derived black hole masses by more than a factor of two.

\subsection{The Dynamical Term}
\label{sec:gdyn}

Our spectral fits assume $g_{\rm dyn} \approx 0$, where $g_{\rm dyn} \simeq -{\rm sgn}(dv^2/dR)v^2/R=g-g_{\rm net}$ is the bulk-motion term of Section~\ref{sec:TLUSTY} that describes any inwards or outwards motion of a layer at radius $R$ and velocity $v$. Here, $g=\frac{G M_\bigstar}{R_\bigstar^2}$ is the physical Newtonian surface gravity of the photosphere with total mass $M_\bigstar$ and radius $R_\bigstar$, while $g_{\rm net}$ is the observed surface gravity. Qualitatively, quasi-stars are stable for $1$--$20$~Myr \citep{Hassan2025,Santarelli2026} while the mass-loss rates that we derive in Section~\ref{sec:disc_popwinds} are much smaller than $M_\bigstar$, so we do not expect this term to be significant. More rigorously, for a quasi-star with a total mass $M_\bigstar$ and an envelope mass $M_{\rm env} \equiv M_\bigstar - M_{\rm BH} \geq 0.67\,M_\bigstar$ that is losing mass at a rate $\dot{M}_{\rm wind}$, the bulk velocity of the photosphere can change by at most $v_{\rm esc}$ over the depletion timescale $t_{\rm dep} = M_{\rm env}/\dot{M}_{\rm wind}$. Using the fact that the dynamical free-fall timescale is $t_{\rm dyn} \approx (R_\bigstar^{3}/GM_\bigstar)^{1/2}\approx \frac{v_{\rm esc}}{\sqrt{2}g}$, we find that $g_{\rm dyn}$ can be at most
\begin{equation}
  \frac{|g_{\rm dyn}|}{g} \leq \frac{v_{\rm esc}\,\dot{M}_{\rm wind}}{g\,M_{\rm env}}
    = \sqrt{2}\,\frac{t_{\rm dyn}}{t_{\rm dep}}.
  \label{eq:gdyn}
\end{equation}
We evaluate Equation~\eqref{eq:gdyn} with the fitted masses and the mass-loss rates of Equation~\eqref{eq:rwind}, finding that, for The Egg, $|g_{\rm dyn}|/g \leq 2\times10^{-2}$, which implies dynamical times of $20$--$200$~yr and depletion times of $\sim$$10^{4}$~yr. Since $M_\bigstar \propto g$, we conclude that $g_{\rm dyn}$ can only shift our masses by at most a few percent. We conclude this section by warning that the actual dynamical situation of the photospheres may be more complex than this, but, as a matter of self-consistency, $g_{\rm dyn}/g\ll1$ is a safe assumption.

\subsection{Non-Zero Angular Momentum}
\label{sec:rotation}

Accretion provides angular momentum at a rate of $f_j\,j_{\rm K}\dot{M}_{\rm acc}$, where $f_j$ is the fraction of the Keplerian value that the arriving gas has, $j_{\rm K} = (GM_\bigstar R_\bigstar)^{1/2}$ is the Keplerian specific angular momentum at the surface, and $\dot{M}_{\rm acc}$ is the rate at which the envelope gains mass. We use $f_j \approx 0.01$ from the $\Omega\Gamma$ limit that \citet{Haemmerle2018} derive, which they describe as the rotation rate at which rotation and radiation pressure counteract gravity. A magnetized wind removes angular momentum at $\tfrac{2}{3}\dot{M}_{\rm wind}\,\Omega R_{\rm A}^{2}$, where $\dot{M}_{\rm wind}$ is the mass-loss rate, $\Omega$ is the angular velocity, and $R_{\rm A}$ is the Alfv\'en radius (i.e., the distance out to which the field forces the wind to corotate; \citealt{udDoula2009}). Conservation then dictates that
\begin{equation}
  f_\Omega \equiv \frac{\Omega}{\Omega_{\rm crit}}
    = \frac{3}{2}\,f_j\,\frac{\dot{M}_{\rm acc}}{\dot{M}_{\rm wind}}
      \left(\frac{R_\bigstar}{R_{\rm A}}\right)^{2},
  \label{eq:spinbalance}
\end{equation}
where $f_\Omega$ is the angular velocity in units of the critical rate $\Omega_{\rm crit} = (GM_\bigstar/R_\bigstar^{3})^{1/2}$.

We evaluate the Alfv\'en radius with $R_{\rm A}/R_\bigstar \approx 0.29 + (\eta_\ast + 1/4)^{1/4}$, where the wind magnetic confinement parameter $\eta_\ast = B_{\rm eq}^{2}R_\bigstar^{2}/(\dot{M}_{\rm wind}v_\infty)$ is the field energy density divided by the wind kinetic energy density, $B_{\rm eq}$ is the equatorial surface field, and $v_\infty$ is the terminal wind speed \citep{udDoula2008, udDoula2009}. $\eta_\ast$ scales as $R_\bigstar^{2}$, where $R_\bigstar\sim10^{16}$~cm, so $\eta_\ast$ reaches $1.4$ to $31$ and $R_{\rm A}/R_\bigstar$ is $1.4$ to $2.7$. In steady state, $f_j = 0.01$ gives $f_\Omega = 0.003$ to $0.009$, and, remains negligible even if we let $f_j$ span plus or minus a decade.

Rotation can cause the equatorial measurement of $g$ to vary by von Zeipel's theorem, i.e., $g_{\rm eff} = g(1 - f_\Omega^2)$ \citep{vonZeipel1924}. The inferred Eddington ratio then becomes,
\begin{equation}
  \Gamma_{\rm es} = \phi\left(1 - f_\Omega^2\right).
  \label{eq:phi_rotation}
\end{equation}
Even in the most conservative cases, this correction can only be a few percent at most.

\subsection{Pressure Balance in a Steady State Atmosphere}
\label{sec:pressurebalance}

We now derive the equilibrium condition of the inflow--outflow interface of Figure~\ref{fig:cartoon}, which we use to estimate how different phenomena affect our mapping between $\phi$ and $\alpha_{\rm acc}$ (Equation~\eqref{eq:alpha_fz}). Energy conservation dictates that
\begin{equation}
  L_{\rm obs} = L_{\rm BH} + L_{\rm in} - \Delta L_{\rm wind},
  \label{eq:lobs}
\end{equation}
where $L_{\rm in} = GM_\bigstar\dot{M}_{\rm in}/R_\bigstar$ is the accretion luminosity of the infall and $\Delta L_{\rm wind} = \tfrac{1}{2}\dot{M}_{\rm wind}(v_\infty^2 + v_{\rm esc}^2)$ is the power that the wind extracts. Dividing the equation by $L_{\rm Edd}(\kappa_{\rm es}) = 4\pi G M_\bigstar c/\kappa_{\rm es}$ turns each term into an Eddington ratio, and we write the corresponding infall and wind terms as $\phi_{\rm in}= \kappa_{\rm es}\dot{M}_{\rm in}/(4\pi R_\bigstar c)$ and $\phi_{\rm out}=\kappa_{\rm es}\dot{M}_{\rm wind}(v_\infty^2 + v_{\rm esc}^2)/(8\pi G M_\bigstar c)$. Bound bulk motion, rotation, magnetic support, and our \texttt{TLUSTY} plane-parallel bias cause us to overpredict the true Eddington ratio as $\Gamma_{\rm es} = \phi\,(1 - f_\Omega^2)(1 - f_B)(1 - g_{\rm dyn}/g)\,10^{-\Delta\log g}$, where $f_\Omega$ is the spin of Appendix~\ref{sec:rotation}, $f_B = P_B/P_{\rm tot}$ is the magnetic support fraction at the photosphere, $g_{\rm dyn}/g \leq 2\times10^{-2}$ is the bound of Appendix~\ref{sec:gdyn}, and $\Delta\log g \geq 0$ is the plane-parallel bias in dex. Solving for $\alpha_{\rm acc}$ then gives
\begin{equation}
  \alpha_{\rm acc}
    = f(Z)\left[
        \phi\left(1-f_\Omega^2\right)\left(1-f_B\right)\left(1-g_{\rm dyn}/g\right)10^{-\Delta\log g}
        - \phi_{\rm in}
        + \phi_{\rm out}
      \right].
  \label{eq:master}
\end{equation}
Table~\ref{tab:mbh_budget} uses this equation to summarize how each of these corrections affects our mass measurements. 

\begin{deluxetable}{lcccccc}
\tablecaption{Black hole mass bounds ($M_{\rm BH} \leq 0.33\,M_\bigstar$, in $M_\odot$) under the corrections of Equation~\eqref{eq:master}. \label{tab:mbh_budget}}
\tablehead{\colhead{Assumption} & \colhead{$\times M_\bigstar$} & \colhead{Egg} & \colhead{UNCOVER} & \colhead{WIDE} & \colhead{{CAPERS}} & \colhead{RS}}
\startdata
baseline & 1.00 & 225 & {549} & {573} & {486} & 4{,}880 \\
$g_{\rm dyn}/g = 0.02$ & 1.02 & 230 & {560} & {584} & {496} & 4{,}977 \\
$f_B = 0.05$ & 1.05 & 236 & {576} & {602} & {510} & 5{,}124 \\
$\Delta\log g = 0.05$ & 1.12 & 252 & {615} & {642} & {544} & 5{,}465 \\
$f_\Omega = 0.6$ & 1.56 & 351 & {856} & {894} & {758} & 7{,}612 \\
all combined & 1.88 & 423 & {1{,}032} & {1{,}077} & {914} & 9{,}174 \\
\enddata
\end{deluxetable}

\onecolumngrid

\section{Additional Spectral Fits}
\renewcommand{\thefigure}{C\arabic{figure}}
\setcounter{figure}{0}
\label{app:spectra}

\begin{figure*}[t!]
\centering
\includegraphics[width=0.6\textwidth]{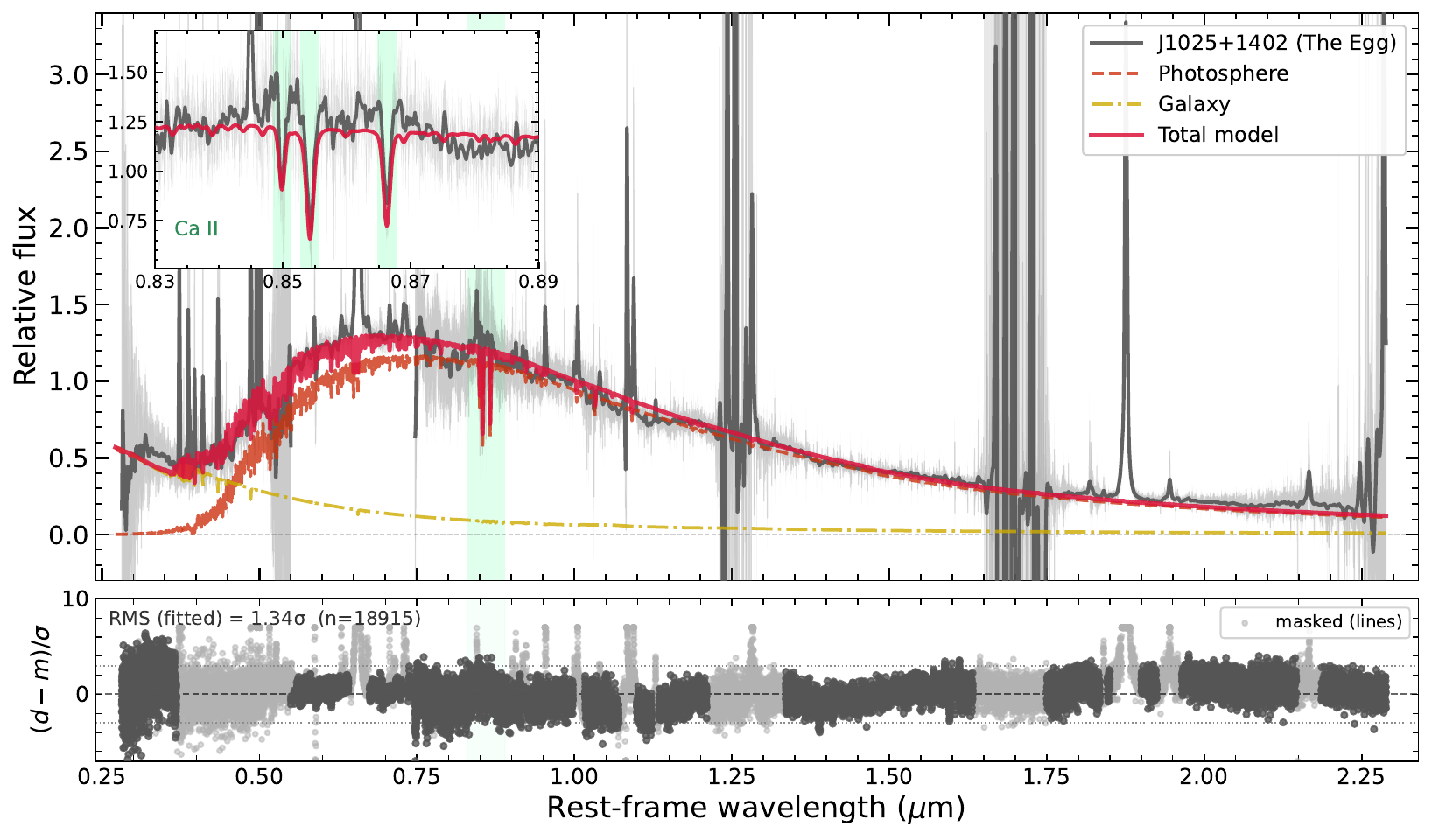}
\caption{Fiducial fit to The Egg (J1025$+$1402, $z = 0.1007$). The dark gray curve is the combined MODS-B, MODS-R, and FIRE spectrum smoothed for display, with the unsmoothed data in light gray and the $1\sigma$ uncertainty band shaded. The red dashed curve is the TLUSTY photosphere ($T = 4502$~K), the gold dash-dotted curve is the host galaxy, and the crimson solid curve is the total model. The green vertical band marks the 0.83--0.89~$\mu$m \ion{Ca}{2} triplet region, which the inset enlarges. The 3 narrow green bands inside the inset mark the individual \ion{Ca}{2} lines.}
\label{fig:egg}
\end{figure*}

\begin{figure*}[t!]
\centering
\includegraphics[width=0.6\textwidth]{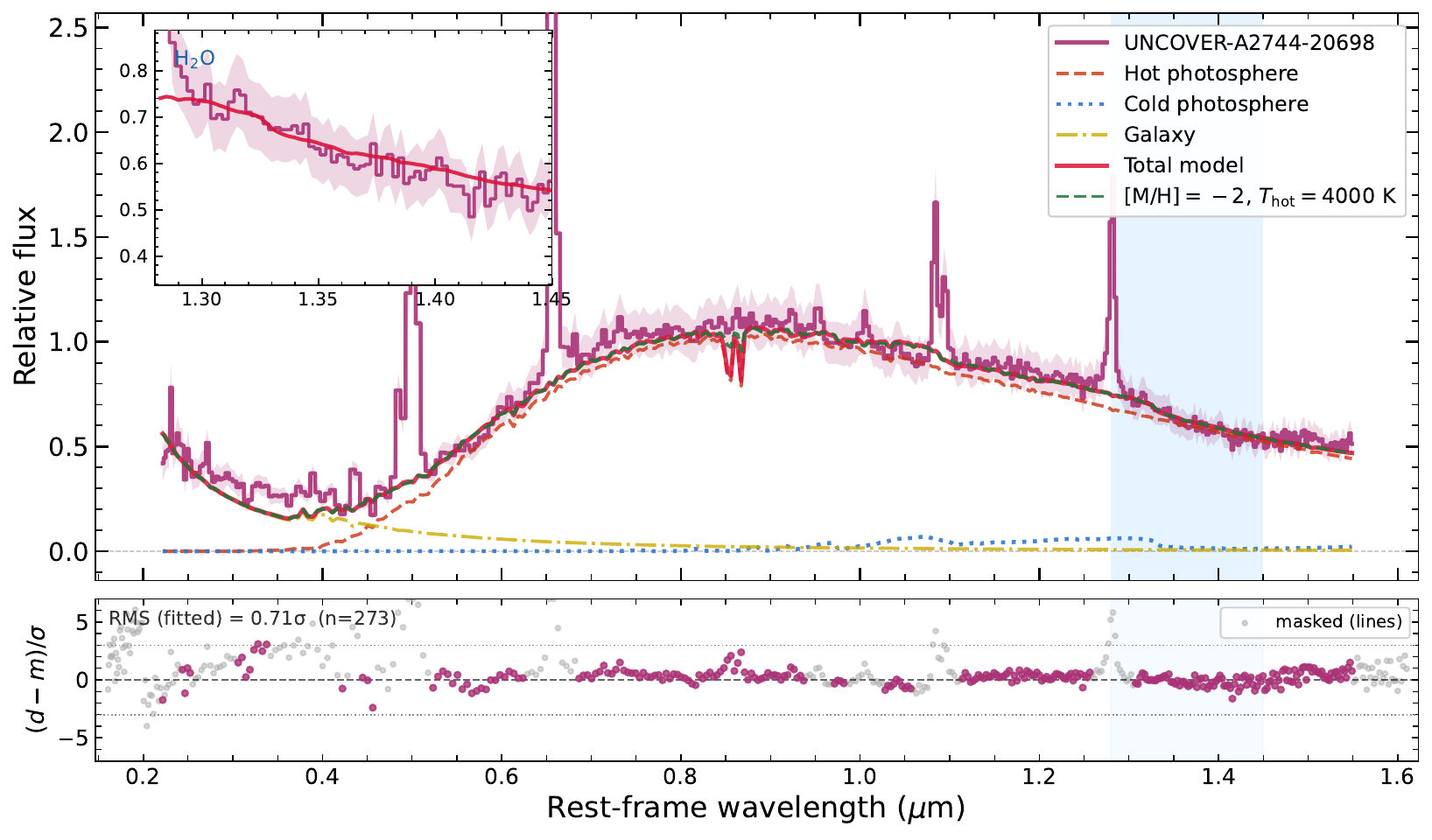}
\caption{Fiducial two-component fit to UNCOVER-A2744-20698 ($z = 2.42$), whose hot and cold photospheres have $T = 3994$~K and $T = {2015}$~K. As in Figure~\ref{fig:wide}, the dark green curve shows the same model with the hot component at $[{\rm M/H}]=-2$.}
\label{fig:uncover}
\end{figure*}

\begin{figure*}[t!]
\centering
\includegraphics[width=0.6\textwidth]{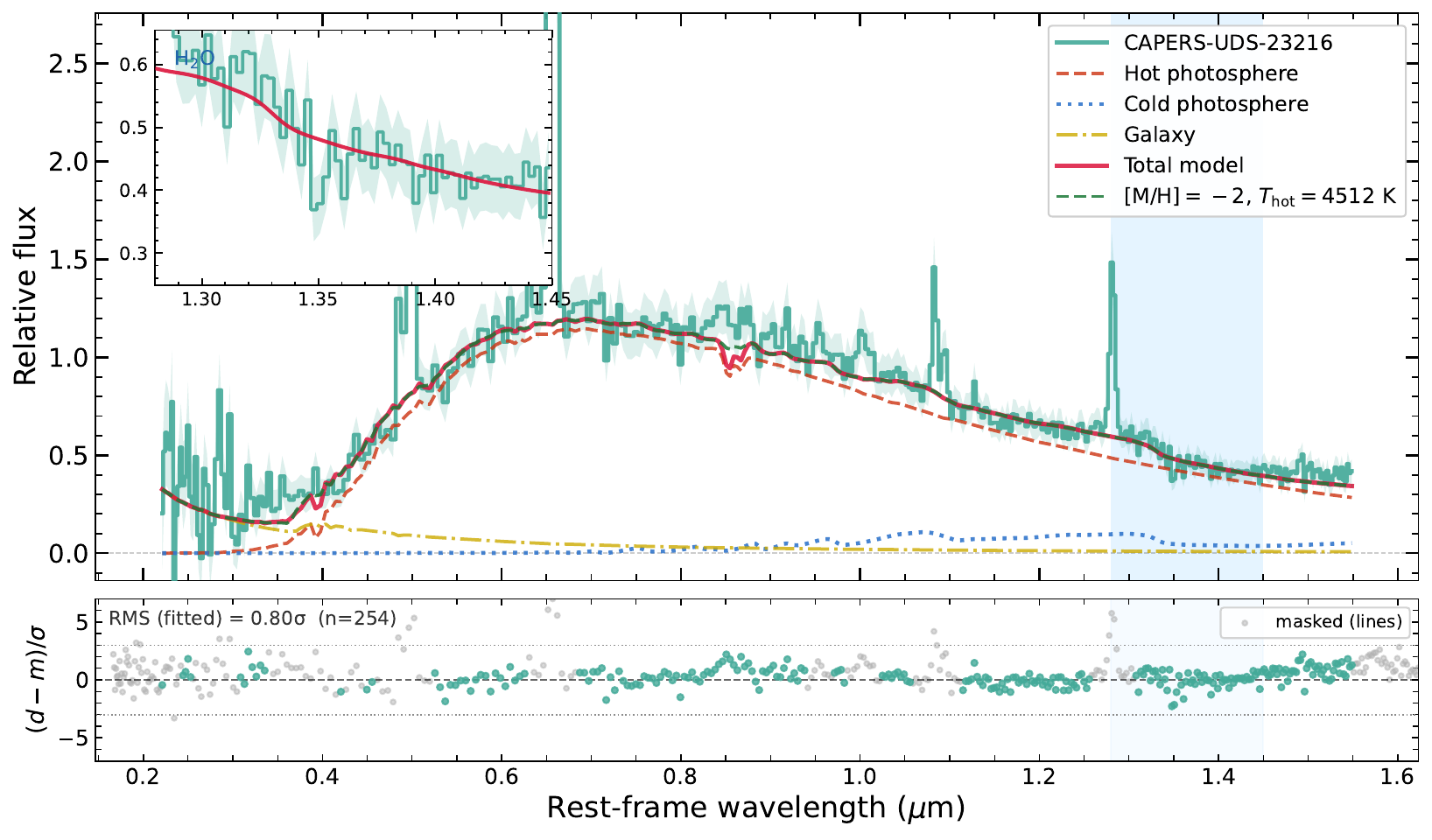}
\caption{Fiducial two-component fit to CAPERS-UDS-23216 ($z = 2.30$), whose hot and cold photospheres have $T = 4512$~K and $T = 2166$~K. As in Figure~\ref{fig:wide}, the dark green curve shows the same model with the hot component at $[{\rm M/H}]=-2$.}
\label{fig:capers}
\end{figure*}

\begin{figure*}[t!]
\centering
\includegraphics[width=0.6\textwidth]{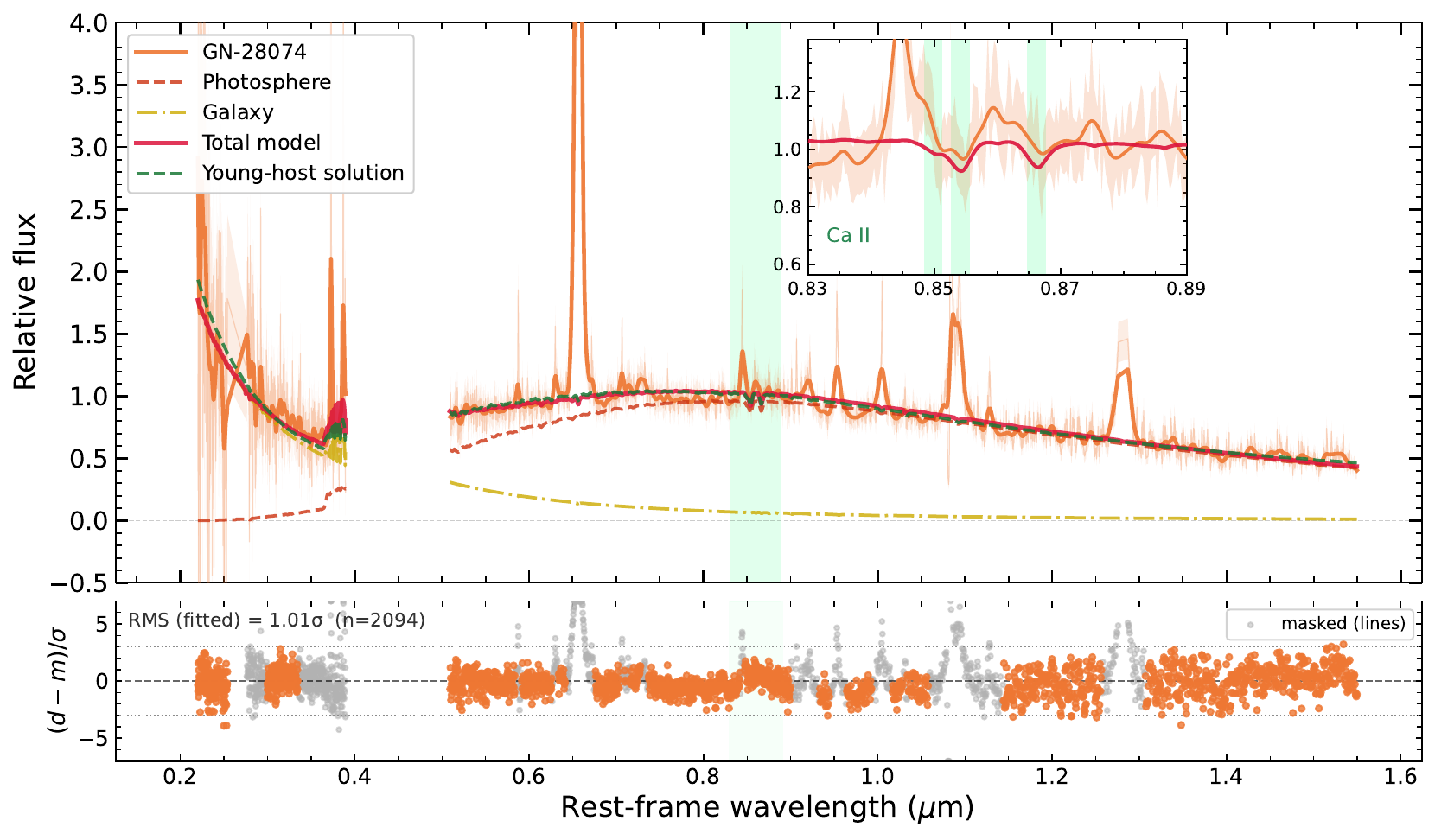}
\caption{Fiducial fit to The Rosetta Stone (GN-28074, $z = 2.26$) across the G140M, G235M, and G395M gratings. The dark orange curve is the spectrum smoothed for display, the light orange line shows the unsmoothed data, and the gap near 0.4--0.5~$\mu$m separates the G140M and G235M wavelength gap. The red dashed curve is the TLUSTY photosphere ($T = 4109$~K), the gold dash-dotted curve is the host galaxy, and the crimson solid curve is the total model. The green vertical band marks the 0.83--0.89~$\mu$m \ion{Ca}{2} triplet region, which the inset enlarges. The dark green curve shows a degenerate, low-$\phi$, low $M_*$ solution that fits the spectrum equally well (see Section~\ref{sec:res_pop}).}
\label{fig:rs}
\end{figure*}

Figures~\ref{fig:egg}, \ref{fig:uncover}, \ref{fig:capers}, and \ref{fig:rs} show our fits to The Egg, UNCOVER-A2744-20698, CAPERS-UDS-23216, and The Rosetta Stone as we describe in Section~\ref{sec:results}.

\bibliography{bibliography}{}
\bibliographystyle{aasjournalv7}

\end{document}